\documentclass[12pt,a4paper]{article}

\usepackage{algorithm}
\usepackage[noend]{algpseudocode}
\usepackage{amsmath}
\usepackage{amssymb}
\usepackage{amsthm}
\usepackage{braket}
\usepackage[subrefformat=parens]{subcaption}
\usepackage{cite}
\usepackage{comment}
\usepackage[nosetpagesize]{graphicx}
\usepackage{here}
\usepackage[setpagesize=false]{hyperref}
\usepackage{latexsym}
\usepackage{mathtools}
\usepackage[nosetpagesize]{xcolor}

\mathtoolsset{showonlyrefs}

\theoremstyle{definition}

\newcommand{\bs}[1]{{\boldsymbol{#1}}}

\usepackage{color}

\def\0{\bs{0}}
\def\1{\bs{1}}
\def\2{\bs{2}}
\def\3{\bs{3}}
\def\4{\bs{4}}
\def\5{\bs{5}}
\def\6{\bs{6}}
\def\7{\bs{7}}
\def\8{\bs{8}}
\def\9{\bs{9}}
\def\a{\bs{a}}
\def\b{\bs{b}}

\def\d{\bs{d}}

\def\f{\bs{f}}

\def\n{\bs{n}}

\def\p{\bs{p}}
\def\q{\bs{q}}
\def\r{\bs{r}}

\def\u{\bs{u}}
\def\v{\bs{v}}

\def\x{\bs{x}}

\def\A{\bs{A}}

\def\I{\bs{I}}
\def\J{\bs{J}}

\def\R{\bs{R}}

\newcommand{\bomega}{\bs\omega}

\newcommand{\norm}[1]{{\left\lVert#1\right\rVert}}
\newcommand{\inv}[1]{{#1^{-1}}}

\newcommand{\figref}[1]{\figurename\ \ref{#1}}

\newcommand{\ang}[1]{\left\langle#1\right\rangle}

\newcommand{\bbR}{\mathbb{R}}

\newcommand{\Rn}[1]{{\bbR^{#1}}}

\makeatletter
\let\@fnsymbol\@arabic
\makeatother

\begin{document}

\date{}

\title{Strongly coupled simulation of incompressible fluid and rigid bodies
	with velocity-based constraints using particle method}
\author{Shugo Miyamoto$^*$\thanks{Department of Systems Innovation, School of Engineering, The University of Tokyo}, Seiichi Koshizuka\footnotemark[1]}

\maketitle

\begin{abstract}
	This paper presents a novel particle method to compute strongly coupled incompressible fluid and rigid bodies. The method adopts a velocity-based formulation and utilizes the linear complementarity problem for the incompressibility constraint. Since all the constraints for incompressibility, inter-rigid-body contacts, and interaction between incompressible fluid and rigid bodies are mathematically compatible, strongly coupled simulation is achieved using the method, where the shapes of the rigid bodies are represented by particles as well. The abstract concept of velocity-based constraints is presented, which generalizes the formulations of the incompressibility constraint and inter-rigid-body contacts and provides a generic way to achieve strongly coupled simulation. Several numerical examples are presented to verify the method, which includes rigid-body computation, hydrostatic pressure, dam-break computation, and circular parch computation for incompressible fluid, buoyancy and seesaw computation for interaction of incompressible fluid and rigid bodies, and complex-scene computation for overall behavior and stability.
\end{abstract}

\section{Introduction}
\label{sec:intro}

Fluid simulation using computers has widely been used for various purposes, such as analysis of industrial products, generating computer graphics, and usage in computer games. Although the dynamics of fluid can be described by the Navier-Stokes equations, solving these equations analytically is unrealistic in many cases and, therefore, solving these equations numerically using computers is important. In reality, fluid interacts with various solids. As such, enabling fluid-solid interaction makes simulation more practical. In addition, within the range of daily handling, many types of fluids and solids can be approximated as incompressible fluids or rigid bodies, respectively. Therefore, coupling of incompressible fluid and rigid body enables more versatile and complex simulation. Coupled simulation of multiple substances can be widely divided into two types of simulation, namely, weakly coupled simulation and strongly coupled simulation. In weakly coupled simulation, interaction between different substances is computed explicitly, whereas, in strongly coupled simulation, this interaction is implicitly computed. Therefore, strongly coupled simulation is more stable and accurate than weakly coupled simulation when the same time step size is adopted.

Rigid-body simulation has long been researched, and impulse-based methods~\cite{mirtich1995impulse} are among the main computation methods. The most important part of rigid-body computation is contact computation, and impulse-based methods deal with contacts in a unified manner by the impulse that integrates force over a small time step. Since simulation requires the integration of time, impulse is easier to handle than force, and using impulse keeps simulation from mathematical and numerical problems~\cite{baraff1991coping, anitescu1997formulating}. If we only consider normal force, we can formulate contact constraints using linear complementarity problems (LCPs)~\cite{baraff1994fast}. When we take friction into consideration, the problem is no longer linear and we have to solve nonlinear complementarity problems (NCPs). However, we can approximate friction to make things less complicated. For example, Tonge et al.~\cite{tonge2012mass} formulated frictional constraints using boxed LCPs by applying pyramid approximation to the Coulomb friction cone. Moreover, Gholami et al.~\cite{gholami2016linear} adopted a continuity approximation to friction to formulate the entire contact problem using LCPs. On the other hand, we can also compute the accurate Coulomb friction cone without approximation by applying an extension to the projected Gauss-Seidel (PGS) method, which is generally used to solve LCPs~\cite{erleben2005physics}. We also use the PGS method and its extension for solving all kinds of constraints that appear in simulation.

When we formulate rigid-body contacts using LCPs, we have to predict the relative velocity after collisions prior to contact computation. This prediction is usually performed using the relationship between the relative velocity and the coefficient of restitution. However, when multiple collisions occur simultaneously, it is difficult to explicitly compute the accurate velocity after these collisions, and LCPs may give physically incorrect results. In addition, depending on the positions and number of contact points, the coefficient matrix of an LCP may not become a symmetric positive-definite matrix, and therefore the solution of an iterative method may not converge to a unique solution~\cite{drumwright2007fast}. Although we do not adopt them in this study, some approaches are presented to solve these problems. For example, Tang et al.~\cite{tang2014impulse} developed the energy tracking impulse (ETI) method, which does not require solving LCPs to compute the impulse. In the ETI method, by tracking energy during collisions, the physically accurate impulse can be calculated without explicitly giving the relative velocity after collision, and its improvement was proposed by Li et al.~\cite{li2020energy} for particle-based rigid-body simulation.

Linear complementarity problems have long been used for rigid-body computation, but there are a few examples of the use of LCPs for fluid computation. Batty et al.~\cite{batty2007fast} used an LCP to formulate wall boundary conditions, and Bodin et al.~\cite{bodin2011constraint} proposed a method by which to simultaneously formulate and solve constant-density constraints and constraints for wall boundary conditions as a mixed LCP. In addition, Gerszewski and Bargteil~\cite{gerszewski2013physics} developed a method by describing the incompressibility of a fluid using inequalities and formulating the resulting pressure equation with an LCP.

Coupled simulation of fluid using a particle method and rigid bodies has long been researched. In the context of the smoothed particle hydrodynamics (SPH) method, Monaghan et al.~\cite{monaghan2003fluid} developed a method that enables coupled simulation of fluid and rigid bodies by representing rigid bodies with particles and computing the interaction force between nearing fluid particles and rigid-body particles. Later, a more stable method was developed by Oh et al.~\cite{oh2009impulse} to compute the interaction based on impulse rather than force, including inter-rigid-body collisions. Akinci et al.~\cite{akinci2012versatile} improved computation at the boundaries of a particle-based rigid-body and fluid and generalized the explicit computation of the interaction so that they can handle the coupled simulation of rigid bodies with fluid computed by both the predictive-corrective SPH method~\cite{solenthaler2009predictive} and the slightly compressible SPH method. Macklin et al.~\cite{macklin2014unified} proposed a method that enables coupled simulation of liquid, gas, solid, and cloth on GPUs by approximating rigid bodies using shape matching techniques. Coupling of rigid bodies and an incompressible fluid has also been researched. Li and Asai~\cite{li2018fluid} proposed a method to couple an incompressible fluid using the incompressible SPH method~\cite{cummins1999sph} and rigid-bodies that are computed using an impulse-based method. There is also a method, which was developed by Klinger et al.~\cite{klingner2006fluid}, to strongly couple an incompressible fluid and rigid bodies using meshes that are dynamically updated during simulation. In the field of computer graphics, incompressible fluid simulation with position-based formulation has been proposed~\cite{bender2017survey}, which can easily be coupled with other position-based methods. However, position-based methods cannot capture some types of physical phenomena, such as variation of coefficient of restitution in rigid-body simulation.

In addition, the moving particle semi-implicit (MPS) method~\cite{koshizuka1996moving} was used to research the coupling of fluids and rigid bodies. Shibata et al.~\cite{shibata2012lagrangian} simulated shipping water on a ship to analyze its effect using weak coupling of fluid and a rigid body. In addition, Shibata et al.~\cite{shibata2013numerical} simulated a lifeboat falling into water also using weak coupling of fluid and a rigid body by transferring fluid pressure to the rigid body. Koshizuka et al.~\cite{koshizuka1998numerical} developed the passively moving solid (PMS) model, which computes rigid bodies with particles by once treating rigid-body particles as fluid particles during pressure and advection computation and then cancelling out the deformation of rigid bodies to restore their shape. The PMS model can handle not only the interaction between rigid bodies and fluid but also between rigid bodies and other rigid bodies. Tanaka et al.~\cite{tanaka2007particle} computed coupled simulation of rigid bodies and an incompressible fluid using a penalty method for inter-rigid-body contact calculation and the PMS model for the interaction of rigid bodies and fluid. Gotoh et al.~\cite{gotoh2006three} computed a flood flow with floating objects using the MPS method and the PMS model for fluid-and-solid coupled simulation. The PMS model itself is not necessarily combined with the MPS method, and the same approach has been used with the SPH method~\cite{bouscasse2013nonlinear} and the weakly compressed SPH method~\cite{ren2015nonlinear} to compute floating rigid bodies. This approach has also been used in the field of computer graphics~\cite{carlson2004rigid}. Large-scale computation of fluid and rigid bodies has been performed by Murotani et al.~\cite{murotani2014development} using the explicit MPS method for fluid and the PMS model for the interaction of fluid and rigid bodies. They divided the computation space into multiple small spaces in order to make the simulation run in parallel~\cite{murotani2014development}. Wang et al.~\cite{wang2019numerical} simulated floating bodies transported by fluid using the MPS method and the PMS model for an incompressible fluid and the interaction of rigid bodies and fluid, as well as the discrete element method for explicitly calculated inter-rigid-body contacts. In the MPS method, the pressure of a fluid is implicitly computed by solving the pressure Poisson equation and thus the pressure field of rigid bodies in the PMS method is implicitly calculated along with that of the fluid when this method is used in combination with the MPS method. As far as we know, there is no other method than ours that achieves fully strongly coupled simulation of incompressible fluid and rigid bodies using a particle method and velocity-based formulations.

Rigid-body computation in impulse-based methods is performed in a velocity-based manner. Therefore, by formulating the incompressibility constraint of fluid in a velocity-based manner as well, we can expect to simultaneously compute both the impulse due to the collision of rigid bodies and the pressure of an incompressible fluid implicitly~\cite{miyamoto2020strong}. In this study, we modify and generalize our previous work~\cite{miyamoto2020strong}. Although the strongly coupled simulation itself has already been achieved in~\cite{miyamoto2020strong}, the core technique of the method highly depends on LCP formulations. In the present paper, we extend the core technique that enables the strongly coupled simulation by introducing the abstract concept of ``velocity-based constraints'' so that it can be applied for wider range. The notion of velocity-based constraints provides a very flexible framework of strongly coupled simulation of various substances that are not limited to fluids and rigid bodies.

The remainder of the present paper is organized as follows. In Section~\ref{sec:fluid}, we introduce the proposed particle method to compute incompressible fluid. We also define the velocity-based constraints in this section. In Section~\ref{sec:rigid}, we describe the impulse-based rigid-body computation and the way to build velocity-based constraints to solve inter-rigid-body collisions. Then, in Section~\ref{sec:rigid-fluid}, the strong coupling of rigid bodies and an incompressible fluid is proposed. We present some numerical examples in Section~\ref{sec:example} in order to confirm the behavior and accuracy of the proposed method. Finally, we present the conclusion in Section~\ref{sec:conclusion}.

\section{Fluid Simulation}
\label{sec:fluid}

In this section, we introduce a velocity-based method to simulate incompressible fluids. We use $\r$, $\u$, $m$, $\rho$, $p$, $\nu$, and $\f$ to represent position, velocity, mass, density, pressure, kinematic viscosity, and acceleration due to external forces, respectively. We use subscript $\phi_i$ to represent arbitrary physical quantity $\phi$ of particle $i$.

\subsection{Governing Equation}
\label{sec:fluid:gov}

We use the Navier-Stokes equation
\begin{align}
	\label{eq:fluid:ns}
	\frac{D\u}{Dt} = -\frac1\rho\nabla p + \nu\nabla^2\u + \f
\end{align}
and the continuity equation
\begin{align}
	\label{eq:fluid:conti}
	\nabla\cdot\u = 0
\end{align}
as the governing equations of incompressible fluids. Here, we assumed that the density and the viscosity are constant. Note that $D\u/Dt$ in the left-hand side of \eqref{eq:fluid:ns} represents the Lagrange derivative of $\u$.

\subsection{Weighting Function}
\label{sec:fluid:weighting}

We denote the weighting function as $w$ and the effective radius as $r_e$. In the proposed method, the derivative of weighting function $w'$ is used during computation, so it is preferable that $w'$ be continuous around effective radius $r_e$. Thus, we use the following function as the weighting function:
\begin{align}
	w(r) = \begin{cases}
		(1-\frac{r}{r_e})^2 & (r<r_e)            \\
		0                   & (\text{otherwise})
	\end{cases}
\end{align}

Using the weighting function, we define the particle number density of particle $i$ as
\begin{align}
	\label{eq:fluid:pnd}
	n_i = \sum_{j\neq i} w(r_{ij}),
\end{align}
where $r_{ij} = \norm{\r_{ij}}$ and $\r_{ij} = \r_j-\r_i$.
We also define the standard particle number density $n^0$ as the particle number density in the initial particle arrangement of a particle that is placed inside a sufficient volume of the fluid. We use $l$ to denote the interval between particles in the initial particle arrangement.

\subsection{Spatial Discretization}
\label{sec:fluid:disc}

We define the discretized gradient model as
\begin{align}
	\label{eq:fluid:grad}
	\ang{\nabla\phi}_i = C \sum_{j\neq i} (\phi_i + \phi_j) w'(r_{ij}) \frac{\r_{ij}}{r_{ij}},
\end{align}
which is widely used in various SPH-based methods, except for the scalar constant $C$. We introduce the constant because the most standard gradient model used in the SPH method tends to give an incorrect value, especially when the effective radius is not large enough. The value of $C$ is chosen so that the gradient model shows the correct value in the initial particle arrangement. More precisely, $C$ is computed as
\begin{align}
	C = \left(\sum_{j\neq i} (\r_i+\r_j)_x  w'(r_{ij}) \frac{(\r_{ij})_x}{r_{ij}}\right)^{-1}
\end{align}
in the initial particle arrangement, where $(\cdot)_x\colon \Rn{3}\to\bbR$ is the first component of a vector and particle $i$ is inside a sufficient volume of the fluid.

Using the gradient model of \eqref{eq:fluid:grad}, the pressure term for particle $i$ is discretized as follows:
\begin{align}
	\label{eq:fluid:pgrad}
	-\frac1\rho\langle\nabla p\rangle_i=-\frac C\rho \sum_{j\neq i} (p_i + p_j) w'(r_{ij}) \frac{\r_{ij}}{r_{ij}}.
\end{align}

\subsection{Incompressibility Constraint}
\label{sec:fluid:incompl}

Since the particle number density is proportional to the fluid density, we can adopt the following equation as the incompressibility condition:
\begin{align}
	\label{eq:fluid:incompl}
	n_i=n^0.
\end{align}
From \eqref{eq:fluid:incompl}, we obtain
\begin{align}
	\label{eq:fluid:nulldiv}
	\frac{dn_i}{dt}=0.
\end{align}
By the definition of the particle number density, we can directly calculate the left-hand side of \eqref{eq:fluid:nulldiv} as:
\begin{align}
	\frac{dn_i}{dt}
	 & = \frac{d}{dt} \sum_{j\neq i} w(r_{ij})                                          \\
	 & = \sum_{j\neq i} w'(r_{ij}) \frac{d}{dt}\norm{\r_j-\r_i}                         \\
	 & = \sum_{j\neq i} w'(r_{ij}) \frac{(\r_j-\r_i)\cdot(\u_j-\u_i)}{\norm{\r_j-\r_i}} \\
	 & = \sum_{j\neq i} w'(r_{ij}) (\u_j-\u_i)\cdot\frac{\r_{ij}}{r_{ij}},
\end{align}
and thus the discretized version of the zero-divergence condition of \eqref{eq:fluid:conti} can be written as:
\begin{align}
	\label{eq:fluid:dconti}
	\sum_{j\neq i} w'(r_{ij}) (\u_j-\u_i)\cdot\frac{\r_{ij}}{r_{ij}} = 0.
\end{align}
However, solving only \eqref{eq:fluid:dconti} as the constraint in actual computation makes the numerical error accumulate as the computation progresses, which allows the fluid to be gradually compressed, and eventually the computation collapses. There are several approaches to avoid this problem. One approach is to utilize the particle shifting techniques \cite{xu2009accuracy, lind2012incompressible} to maintain the volume of the fluid constant. Another approach is to mix the positional constraint violation term to the velocity constraint. The latter is easier to handle because this approach requires few changes to the entire procedure and keeps the equations to be solved linear. We use this approach to solve the problem. Let $\alpha\in[0,1]$ be a constant to control the amount of mixing. We use the following equation to obtain the constraint:
\begin{align}
	\label{eq:fluid:mixed}
	\sum_{j\neq i} w'(r_{ij}) (\u_j-\u_i)\cdot\frac{\r_{ij}}{r_{ij}}
	 & = \frac\alpha h (n^0 - n_i),
\end{align}
where $h$ is the time step size. Setting $\alpha$ to zero means that there is no positional term and \eqref{eq:fluid:mixed} is equal to \eqref{eq:fluid:dconti}, whereas setting $\alpha$ to one means that an attempt is made to cancel out all positional constraint violations at the next time step. However, using a large value of $\alpha$ does not work well due to the linear approximation and makes the simulation unstable, so we use 0.05 for the value of $\alpha$ throughout the present paper.

\subsection{Time Integration}
\label{sec:fluid:int}

In this method, we use a semi-implicit scheme for the time integration, which is similar to the MPS method. The computation in each time step is performed roughly as follows. We first calculate the viscosity and the external force terms explicitly to obtain temporal velocity $\u^*$, and then we compute the pressure to correct the temporal velocity so that $\u^*$ satisfies constraint \ {eq:fluid:mixed}. Finally, we adopt the corrected temporal velocity as the velocity in the next time step, and update the position using the velocity.

We use integer $t$ to denote a time step, and use superscript $\phi^t$ to denote arbitrary physical quantity $\phi$ at time $t$. We do not define the detailed algorithm for computing explicit terms, as there are many acceptable ways to compute these terms and the subsequent procedure is not affected as long as we can obtain temporal velocity $\u^*$. In the following discussion, we focus on the computation of $\u^{t+1}$ and $p^{t+1}$.

Let $\x^t$ and $\u^t$ be the position and the velocity of particles, respectively, at time $t$. After obtaining temporal velocity $\u^*$ by explicit calculation, we first initialize pressure $p^{t+1}$ with zero. Pressure $p^{t+1}$ plays a role in correcting temporal velocity $\u^*$ through the pressure term $-\frac1\rho\ang{p^{t+1}}$. In order to derive an equation to obtain $p^{t+1}$, assume that $\u^{t+1} = \u^* - \frac h\rho\ang{p^{t+1}}$ holds and assign $\u^{t+1}$ to constraint \eqref{eq:fluid:mixed} to obtain
\begin{align}
	\frac\alpha h (n^0 - n_i)
	 & = \sum_{j\neq i} w'(r_{ij}) (\u_j^{t+1}-\u_i^{t+1})\cdot\frac{\r_{ij}}{r_{ij}}                                                           \\
	 & = \sum_{j\neq i} w'(r_{ij}) \left(\u_j^*-\u_i^*-\frac h\rho\ang{p_j^{t+1}}+\frac h\rho\ang{p_i^{t+1}}\right)\cdot\frac{\r_{ij}}{r_{ij}},
\end{align}
or, equivalently,
\begin{align}
	 & \frac\alpha h (n^0 - n_i) - \sum_{j\neq i} w'(r_{ij}) (\u_j^*-\u_i^*)\cdot\frac{\r_{ij}}{r_{ij}} \\
	\label{eq:fluid:eqlast}
	 & = h\frac C\rho\sum_{j\neq i} w'(r_{ij}) \left(
	\sum_{k\neq i} (p_i + p_k) w'(r_{ik}) \frac{\r_{ik}}{r_{ik}}
	-\sum_{k\neq j} (p_j + p_k) w'(r_{jk}) \frac{\r_{jk}}{r_{jk}}
	\right)\cdot\frac{\r_{ij}}{r_{ij}}.
\end{align}
Let $N$ be the number of particles; $\b\in\Rn N$ be a vector, the $i$th component of which is $\frac\alpha h (n^0 - n_i) - \sum_{j\neq i} w'(r_{ij}) (\u_j^*-\u_i^*)\cdot\frac{\r_{ij}}{r_{ij}}$; and $\p\in\Rn N$ be a vector, the $i$th component of which is $p_i^{n+1}$. Then, there exists a matrix $\A\in\Rn{N\times N}$ such that $\A\p=\b$ is equivalent to equation \eqref{eq:fluid:eqlast}. Define $\v_{ij}\in\Rn3$ as
\begin{align}
	\v_{ij} = \begin{cases}
		\0                                & (i=j)      \\
		w'(r_{ij})\dfrac{\r_{ij}}{r_{ij}} & (i\neq j).
	\end{cases}
\end{align}
Then, the right-hand side of \eqref{eq:fluid:eqlast} can be written as
\begin{align}
	 & \phantom{=} h\frac C\rho\sum_{j\neq i}\sum_{k\neq i}(p_i+p_k)w'(r_{ij})\frac{\r_{ij}}{r_{ij}}\cdot w'(r_{ik})\frac{\r_{ik}}{r_{ik}}  \\
	 & \phantom{=} -h\frac C\rho\sum_{j\neq i}\sum_{k\neq j}(p_j+p_k)w'(r_{ij})\frac{\r_{ij}}{r_{ij}}\cdot w'(r_{jk})\frac{\r_{jk}}{r_{jk}} \\
	 & = h\frac C\rho\sum_{j,k}(p_i+p_k)\v_{ij}\cdot\v_{ik} - h\frac C\rho\sum_{j,k}(p_j+p_k)\v_{ij}\cdot\v_{jk}                            \\
	 & = h\frac C\rho\biggl(p_i\sum_{j,k}\v_{ij}\cdot\v_{ik} + \sum_jp_j\sum_k\v_{ij}\cdot\v_{ik}                                           \\
	 & \phantom{=} -\sum_jp_j\sum_k\v_{ij}\cdot\v_{jk} + \sum_jp_j\sum_k\v_{ik}\cdot\v_{jk}\biggr)                                          \\
	 & = p_ih\frac C\rho\left(\norm{\sum_k\v_{ik}}^2+\sum_k\norm{\v_{ik}}^2\right)                                                          \\
	 & \phantom{=} +\sum_{j\neq i}p_jh\frac C\rho\sum_k\left(\v_{ij}\cdot\v_{ik}-\v_{ij}\cdot\v_{jk}+\v_{ik}\cdot\v_{jk}\right).
\end{align}
Therefore, the components of matrix $\A$ are
\begin{align}
	A_{ij} = h\frac C\rho\begin{dcases}
		\norm{\sum_k\v_{ik}}^2+\sum_k\norm{\v_{ik}}^2                                  & (i=j)      \\
		\sum_k\left(\v_{ij}\cdot\v_{ik}-\v_{ij}\cdot\v_{jk}+\v_{ik}\cdot\v_{jk}\right) & (i\neq j).
	\end{dcases}
\end{align}
Let $\J_i\in\Rn{3N}$ be a vector
\begin{align}
	\J_i = \begin{pmatrix}
		\v_{i,1}        \\
		\vdots          \\
		\v_{i,i-1}      \\
		-\sum_j\v_{i,j} \\
		\v_{i,i+1}      \\
		\vdots          \\
		\v_{i,N}
	\end{pmatrix},
\end{align}
and $\J\in\Rn{3N\times N}$ be a matrix $\J=(\J_1 \cdots \J_N)$. Then, matrix $\A$ can be decomposed into $\A=h\frac C\rho\J^T\J$. Indeed, the inner product of $\J_i$ and $\J_j$ times $h\frac C\rho$ is
\begin{align}
	h\frac C\rho\J_i\cdot\J_j
	 & = h\frac C\rho\begin{dcases}
		\sum_k\v_{ik}\cdot\v_{ik} + (\sum_k\v_{ik})\cdot(\sum_k\v_{ik})                              & (i=j)     \\
		\sum_{k\neq i,j} \v_{ik}\cdot\v_{jk} - \v_{ij}\cdot\sum_k\v_{jk} - \v_{ji}\cdot\sum_k\v_{ik} & (i\neq j)
	\end{dcases} \\
	 & = h\frac C\rho\begin{dcases}
		\sum_k\norm{\v_{ik}}^2 + \norm{\sum_k\v_{ik}}^2                       & (i=j)     \\
		\sum_k(\v_{ik}\cdot\v_{jk} - \v_{ij}\cdot\v_{jk}+\v_{ij}\cdot\v_{ik}) & (i\neq j)
	\end{dcases} \\
	 & = A_{ij},
\end{align}
and, therefore, if the particles are well arranged, in other words, if matrix $\J$ is not degenerated, then matrix $\A$ is symmetric positive definite.

As discussed above, we need to solve the equation $\A\p=\b$ with respect to $\p$ in order to correct the temporal velocity so that constraint \eqref{eq:fluid:mixed} is satisfied. However, directly using the pressure obtained by solving the equation may make the simulation unstable. Particle methods sometimes behave poorly when there is an attractive force between particles. Considering the gradient model of \eqref{eq:fluid:grad}, we see that the attractive force due to the pressure works between particles $i$ and $j$ if and only if $p_i+p_j<0$ holds for some $i$ and $j$. This fact leads us to an easy sufficient condition: there is no attractive force if $p_i\geq0$ holds for every particle $i$. The simplest way to make the pressure nonnegative is just to project the solution of $\A\x=\b$ onto $\bbR_+^N$, where $\bbR_+=[0,\infty)$. However, this modification of the solution breaks the equilibrium between the pressure of each particle. If the pressure of particle $i$ is changed from negative to zero, then the $j$th row of equation $\A\p=\b$ will no longer hold for each particle $j$ that is a neighboring particle of particle $i$. We need $\p$ to satisfy $\A\p=\b$ as much as possible, whereas $\p$ is forced to be nonnegative. These requirements can be satisfied by finding $\p$ that satisfies
\begin{align}
	\label{eq:fluid:lcp1}
	(\p)_i            & \geq0 \\
	\label{eq:fluid:lcp2}
	(\A\p-\b)_i       & \geq0 \\
	\label{eq:fluid:lcp3}
	(\p)_i(\A\p-\b)_i & =0
\end{align}
for each $i$, where $(\cdot)_i$ is the $i$th component of a vector.
Inequality \eqref{eq:fluid:lcp1} is for the nonnegativity of $\p$. Inequality \eqref{eq:fluid:lcp2} is for the requirement that $\A\p=\b$ should hold as much as possible. If $(\A\p-\b)_i<0$ holds, then we can make $(\A\p-\b)_i=0$ hold by increasing $(\p)_i$, because $\A$ is positive definite and, therefore, $A_{ii}>0$. Equation \eqref{eq:fluid:lcp3} has the same requirement that $\A\p=\b$ should hold as much as possible. If $(\p)_i(\A\p-\b)_i>0$ holds, then $(\p)_i$ is unnecessarily large, and either $(\p)_i=0$ or $(\A\p-\b)_i=0$ can be achieved by decreasing $(\p)_i$ because $A_{ii}>0$ holds. The problem of finding $\p$ that satisfies \eqref{eq:fluid:lcp1}, \eqref{eq:fluid:lcp2}, and \eqref{eq:fluid:lcp3} is called a linear complementarity problem (LCP). An LCP has a unique solution if the coefficient matrix of the problem is symmetric positive definite. Since $\A$ is proven to be symmetric positive definite, there exists a unique solution of the problem.

We configured matrix $\A$ in order to define the problem. However, in actual computation, we do not compute the matrix explicitly. Instead, we use temporal velocity $\u^*$ to iteratively solve this model. Since the constraint is originally imposed on the velocity, we repeatedly apply the change in pressure to the temporal velocity using the gradient model of \eqref{eq:fluid:pgrad}, while determining the change in pressure by computing the current deviation of $\u^*$ from constraint \eqref{eq:fluid:mixed}. More precisely, we iterate the following steps until some stopping criteria are satisfied. For each particle $i$, we first compute the current deviation of constraint $\delta_i$ as
\begin{align}
	\delta_i = \frac\alpha h (n^0 - n_i) - \sum_{j\neq i} w'(r_{ij}) (\u_j^*-\u_i^*)\cdot\frac{\r_{ij}}{r_{ij}}.
\end{align}
Then, we compute the next value of pressure $p'_i$ as
\begin{align}
	p'_i = \max\left\{0,\ p_i^{t+1} + \frac{\delta_i}{A_{ii}}\right\}.
\end{align}
Finally, we update pressure $p_i^{t+1}$ and apply the difference of it to temporal velocity $\u^*$:
\begin{align}
	\Delta p_i & = p'_i - p_i^{t+1}                                                                      \\
	p_i^{t+1}  & \gets p'_i                                                                              \\
	\u_i^*     & \gets \u_i^* - h\frac C\rho \sum_{j\neq i} \Delta p_i w'(r_{ij}) \frac{\r_{ij}}{r_{ij}} \\
	\u_j^*     & \gets \u_j^* + h\frac C\rho \Delta p_i w'(r_{ij}) \frac{\r_{ij}}{r_{ij}}\quad(j\neq i).
\end{align}
This iterative approach actually corresponds to the projected Gauss-Seidel (PGS) method, which is an iterative method to solve LCPs. The tentative solution is updated repeatedly in a Gauss-Seidel manner and is projected onto the nonnegative space.

Remarkably, we can compute pressure $p^{t+1}$ based only on temporal velocity $\u^*$. When we update the pressure of a particle, we no longer need to refer the pressure of the neighboring particles because that information can be obtained through the temporal velocity of the neighboring particles. This makes the method very flexible because we can actually apply external force and change the temporal velocity in the middle of the iteration process. When this happens, the pressure starts adapting to the new environment and finally converges to a different distribution from that which was previously assumed. This enables us to strongly couple the fluid simulation with different types of simulation, as long as these simulations only have velocity-based constraints.

We define velocity-based constraints here. A constraint in a system is referred to as velocity-based if it only requires the temporal velocity of the system to iteratively update its constraint force or impulse, and we apply the difference of it  to update the temporal velocity. A velocity-based constraint must not require knowledge of the information on other constraints that cannot be obtained through the temporal velocity of the system. By this definition, we can say that each particle $i$ has a velocity-based constraint, the constraint force of which is $p_i^{t+1}$. We can update the constraint force and reapply this force to the temporal velocity without knowing the pressure of the other particles explicitly.

When the iteration process is terminated, we adopt the temporal velocity as the velocity in the next time step by $\u_i^{t+1}=\u_i^*$ for each particle $i$. Then, we update the position of each particle $i$ by $\r_i^{t+1}=\r_i^t+h\u_i^{t+1}$ to finish the time step.

\subsection{Smoothing Pressure}

As we will show in Section \ref{sec:example}, the computed value of raw pressure $p$ contains a lot of noise, even though the fluid as a whole behaves very smoothly. However, we can actually obtain a smooth and sufficiently accurate pressure field by smoothing $p$ to some extent. A similar approach is presented by Kondo~\cite{kondo2020physically} to decrease the heavy noise of pressure. In \cite{kondo2020physically}, the virial theorem is used to make the smoothing process physically meaningful. Considering the balance of smoothness and locality, we define smoothed pressure $\tilde p_i$ for each particle $i$ by:
\begin{align}
	\label{eq:fluid:spressure}
	\tilde p_i = \frac{\sum_j w_\text{smooth}(r_{ij})p_j}{\sum_j w_\text{smooth}(r_{ij})},
\end{align}
which is equal to the weighted average of $p_i$ with weighting function $w_\text{smooth}$. We use the following weighting function for smoothing:
\begin{align}
	w_\text{smooth}(r) = \begin{cases}
		(r_e^2-r^2)^3 & (r<r_e)             \\
		0             & (\text{otherwise}),
	\end{cases}
\end{align}
which is proportional to a standard weighting function in the SPH method and has the same effective radius as the weighting function used in the main fluid computation.

\section{Rigid-body Simulation}
\label{sec:rigid}

In this section, we introduce a rigid-body simulation method that can be strongly coupled with the incompressible fluid simulation introduced in the last section. As mentioned previously, in order to enable strong coupling, the rigid-body simulation must only have velocity-based constraints. This requirement, however, is not hard to satisfy because most of the constraints imposed on rigid bodies can be written as velocity-based constraints, which we defined in the last section.

We use $\x$, $\v$, $\bomega$, $m$, $\I$, $\R$, and $\q$ to represent a the center of gravity, linear velocity, angular velocity, mass, moment of inertia, rotation matrix, and a unit quaternion that represents the orientation, respectively, of a rigid body.

\subsection{Time Integration}
\label{sec:rigid:time}

Let $h$ be the time step size. We use the same notation as in Section \ref{sec:fluid}. We use $\phi^t$ to denote physical quantity $\phi$ at time $t$. We integrate the center of gravity $\x$ of a rigid body as follows:
\begin{align}
	\x^{t+1} = \x^t + h\v^{t+1}.
\end{align}
In addition, we integrate quaternion $\q$ of a rigid body as follows:
\begin{align}
	\label{eq:rigid:qint}
	\q^{t+1} = \d\q^{t+1}\q^t,
\end{align}
where $\d\q^{t+1}$ is a unit quaternion that represents a rotation, the rotation axis of which is $\bomega^{t+1}$ and the rotation angle of which is $h\norm{\bomega^{t+1}}$. Since the moment of inertia tensor $\I$ depends on the orientation of the rigid body, using only \eqref{eq:rigid:qint} does not accurately conserve angular momentum $\I\bomega$. However, computing an accurate angular velocity requires implicit calculation due to a numerical instability and is therefore costly. In the present paper, we assume that a change in angular velocity only occurs due to an external force applied to the rigid body. If the conservation of the angular momentum is important, a gyroscopic force can be calculated as an external force, which can be applied to a rigid body.

The linear velocity and the angular velocity at time $t+1$ is computed as follows. Starting from the previous velocities $\v^t$ and $\bomega^t$, we first compute temporal velocities $\v^*$ and $\bomega^*$ by applying external forces explicitly and then correct velocities by applying constraint impulses iteratively, so that they  satisfy all of the velocity-based constraints. Finally, we adopt the corrected temporal velocities as the velocities at time $t+1$ as $\v^{t+1}=\v^*$ and $\bomega^{t+1}=\bomega^*$.

\subsection{Contact Constraint}
\label{sec:rigid:contact}

When two rigid bodies touch or collide, we generate contact points between these bodies and handle these bodies as contact constraints. Contact points are usually generated as they form the convex hull of the touching area. In the present paper, each contact point is treated as an independent constraint, and we compute the constraint impulse of each contact point independently.

Consider that two rigid bodies, namely, $A$ and $B$, are touching at point $\p$ with the unit contact normal $\n$, where the contact normal is directed from rigid body $B$ to rigid body $A$ at the contact point. We use subscript $\phi_X$ to denote arbitrary physical quantity $\phi$ of rigid body $X$. Let $\r_A=\p-\x_A$ and $\r_B=\p-\x_B$ be the relative contact positions, and let $N$ be the contact impulse along $\n$. Then, impulses $N\n$ and $-N\n$ are applied to rigid bodies $A$ and $B$, respectively. The changes in velocities of each rigid body caused by the impulses are as follows:
\begin{align}
	\Delta\v_A      & = \frac1{m_A}N\n               \\
	\Delta\bomega_A & = \inv{I_A} (\r_A\times N\n)   \\
	\Delta\v_B      & = -\frac1{m_B}N\n              \\
	\Delta\bomega_B & = -\inv{I_B} (\r_B\times N\n).
\end{align}
We define the relative velocity at contact point $\v_{AB}$ by
\begin{align}
	\v_{AB} = (\v_A+\bomega_A\times\r_A) - (\v_B+\bomega_B\times\r_B).
\end{align}
Then, the change in relative velocity $\Delta\v_{AB}$ can be written as
\begin{align}
	\Delta\v_{AB} & = (\Delta\v_A+\Delta\bomega_A\times\r_A) - (\Delta\v_B+\Delta\bomega_B\times\r_B)                                          \\
	              & = \left(\frac1{m_A}N\n+(\inv{I_A} (\r_A\times N\n))\times\r_A\right)                                                       \\
	              & \phantom{=} + \left(\frac1{m_B}N\n+(\inv{I_B} (\r_B\times N\n))\times\r_B\right)                                           \\
	              & = \left(\frac1{m_A}+\frac1{m_B}\right)N\n - [\r_A\times](\inv{I_A} ([\r_A\times]N\n))                                      \\
	              & \phantom{=} - [\r_B\times](\inv{I_B} ([\r_B\times]N\n))                                                                    \\
	\label{eq:rigid:drv}
	              & = \left(\frac1{m_A}\1+\frac1{m_B}\1 + [\r_A\times]^T\inv{I_A}[\r_A\times] + [\r_B\times]^T\inv{I_B}[\r_B\times]\right)N\n,
\end{align}
where $[\a\times]$ is a skew-symmetric matrix that satisfies $[\a\times]\b=\a\times\b$ for arbitrary vectors $\a$ and $\b$, and $\1$ is the identity matrix of order three. By taking the dot product of \eqref{eq:rigid:drv} and $\n$, we obtain the change in relative velocity along the normal:
\begin{align}
	\Delta\v_{AB}\cdot\n
	 & = \n^T\left(\frac1{m_A}\1+\frac1{m_B}\1 + [\r_A\times]^T\inv{I_A}[\r_A\times] + [\r_B\times]^T\inv{I_B}[\r_B\times]\right)N\n \\
	 & = \left(\frac1{m_A}+\frac1{m_B} + \norm{\r_A\times\n}_\inv{\I_A}^2 + \norm{\r_B\times\n}_\inv{\I_B}^2\right)N                 \\
	\label{eq:rigid:drvn}
	 & = \inv{m_{AB\n}} N.
\end{align}
Here, $\norm{\a}_{\A}=\sqrt{\a^T\A\a}$ for an arbitrary vector $\a$ and symmetric positive definite matrix $\A$, and the effective mass along normal $m_{AB\n}$ is defined as
\begin{align}
	m_{AB\n} = \left(\frac1{m_A}+\frac1{m_B} + \norm{\r_A\times\n}_\inv{\I_A}^2 + \norm{\r_B\times\n}_\inv{\I_B}^2\right)^{-1}.
\end{align}
From \eqref{eq:rigid:drvn}, we obtain the linear relationship between $\Delta\v_{AB}\cdot\n$ and $N$. In order to change $\v_{AB}\cdot\n$ by some amount $\varepsilon$, we need to set $N$ to $m_{AB\n}\varepsilon$. Note that in actual computation, we iteratively update the value of $N$ as well as the temporal velocities of the rigid bodies.

The constraint of the contact point can be written as follows in the form of an LCP:
\begin{align}
	\v_{AB}^*\cdot\n - b_{AB}    & \geq 0 \\
	N                            & \geq 0 \\
	N(\v_{AB}^*\cdot\n - b_{AB}) & = 0,
\end{align}
where $b_{AB}$ is the target velocity of the contact point. The target velocity represents how fast the rigid bodies move away from each other along the normal after the collision. Using the coefficient of restitution $e$, we can estimate the relative velocity along the normal after the collision as $-e\v_{AB}^*\cdot\n$. Note that this value must be observed and fixed before any constraint impulse is applied. Although this estimation is not always physically correct because the energy passing through multiple bodies is not tracked, the estimation gives satisfactory results in most cases. Newton's cradle is a typical case in which this estimation does not work well. Since the estimation depends only on the velocities, simply setting $b_{AB}=-e\v_{AB}^*\cdot\n$ will let rigid bodies gradually pass through each other due to numerical error if we do not correct positions separately. This is similar to the case of the incompressibility of fluid shown in Section \ref{sec:fluid:int}. We should either correct positions or mix a positional term with the constraint in order to maintain the incompressibility. We also mix a positional term with the constraint in the case of the rigid-body simulation. Let $d_{AB}$ be the penetration depth of rigid bodies $A$ and $B$ at the contact point, and let $\alpha\in[0,1]$ be a constant to control the amount of mixing. Then, we set $b_{AB}$ as
\begin{align}
	\label{eq:rigid:mixed}
	b_{AB} = \max\{-e\v_{AB}^*\cdot\n, \frac\alpha hd_{AB}\}.
\end{align}

Contact constraints are velocity-based constraints, and we can solve these constraints in the same way as in Section \ref{sec:fluid:int}. We iterate the following steps for each contact constraint, until some stopping criteria are met. First, we compute the difference $\delta_{AB}$ of the target velocity and the current relative velocity along the normal as
\begin{align}
	\delta_{AB} = b_{AB} - \v^*_{AB}\cdot\n.
\end{align}
Then, we compute the next value of impulse $N'$ by
\begin{align}
	N' = \max\{0, N + m_{AB\n}\delta_{AB}\}.
\end{align}
Finally, we update impulse $N$ and apply its difference to each rigid body to update the temporal velocities:
\begin{align}
	\Delta N    & = N' - N                                              \\
	N           & \gets N'                                              \\
	\v^*_A      & \gets \v^*_A + \frac1{m_A}\Delta N\n                  \\
	\bomega^*_A & \gets \bomega^*_A + \inv{I_A}(\r_A\times \Delta N\n)  \\
	\v^*_B      & \gets \v^*_B - \frac1{m_B}\Delta N\n                  \\
	\bomega^*_B & \gets \bomega^*_B - \inv{I_B}(\r_B\times \Delta N\n).
\end{align}

\section{Rigid Body and Fluid Interaction}
\label{sec:rigid-fluid}

We introduced the incompressible fluid computation in Section \ref{sec:fluid}, and the rigid-body computation in Section \ref{sec:rigid}. In this section, we introduce the strongly coupled simulation method of rigid bodies and fluid.

We defined velocity-based constraints that can be strongly coupled with other velocity-based constraints. Thus, for the strongly coupled simulation, we should define the rigid-body and fluid interaction model, which consists of velocity-based constraints.

\subsection{Shape Representation of a Rigid Body}

In the present paper, we adopt particles to represent the shape of a rigid body. Since particles are used to represent fluid, there are many benefits to using particles as rigid bodies.

To construct a rigid body with particles, we first fill the inside of the rigid bodies with particles. Each particle is placed at the center of a grid of interval $l$, which is the same value as the interval of initial particle arrangement for fluid. Then, we compute the mass, the center of gravity, and the inertia tensor matrix of the rigid body. In this step, particles are considered to be axis-aligned boxes of edge length $l$. The position of each particle relative to the center of gravity of the rigid body is computed and stored in order to update the position of the particle when the position and orientation of the rigid body are changed.

During collision detection between rigid bodies, each particle is treated as a sphere of radius $l/2$. In addition, contact information is generated based on the sphere shape. In other words, when a pair of particles collides, a contact point is generated at the middle point between two particles and the contact normal is set to a vector parallel to the relative position vector of two particles. The penetration depth of the contact point is set to $l-r$, where $r$ is the distance between two particles.

In order to compute the particle number density of fluid particles around the boundary of a rigid body and fluid accurately, we include rigid-body particles in the particle number density computation of fluid particles. The particle number density of a fluid particle is computed using all kinds of particles.

\subsection{Contact Constraint between a Rigid Body and a Fluid Particle}

Since rigid bodies and fluid are both represented as particles, contact computation between rigid bodies and fluid can be performed in a particle-based manner. We treat fluid particles as rigid-body particles during the rigid-body and fluid interaction computation. Each fluid particle is considered as a rigid-body particle, which is an independent rigid body. The physical properties of the rigid body are defined based on a sphere of radius $l/2$ and mass $\rho l^d$, where $\rho$ is the density of the fluid and $d$ is the dimension of the simulation. The sphere is placed at the same position as the particle, and the rigid body shares its linear velocity with the velocity of the particle. That means that the linear velocity of the rigid body actually refers to the velocity of the particle, and if one velocity is changed, then the other velocity is also changed to the same value. This relationship is also applied to the temporal linear velocity of the rigid body and the temporal velocity of the particle while the iteration for the constraint resolution is running.

\begin{figure}
	\centering
	\includegraphics[width=\textwidth]{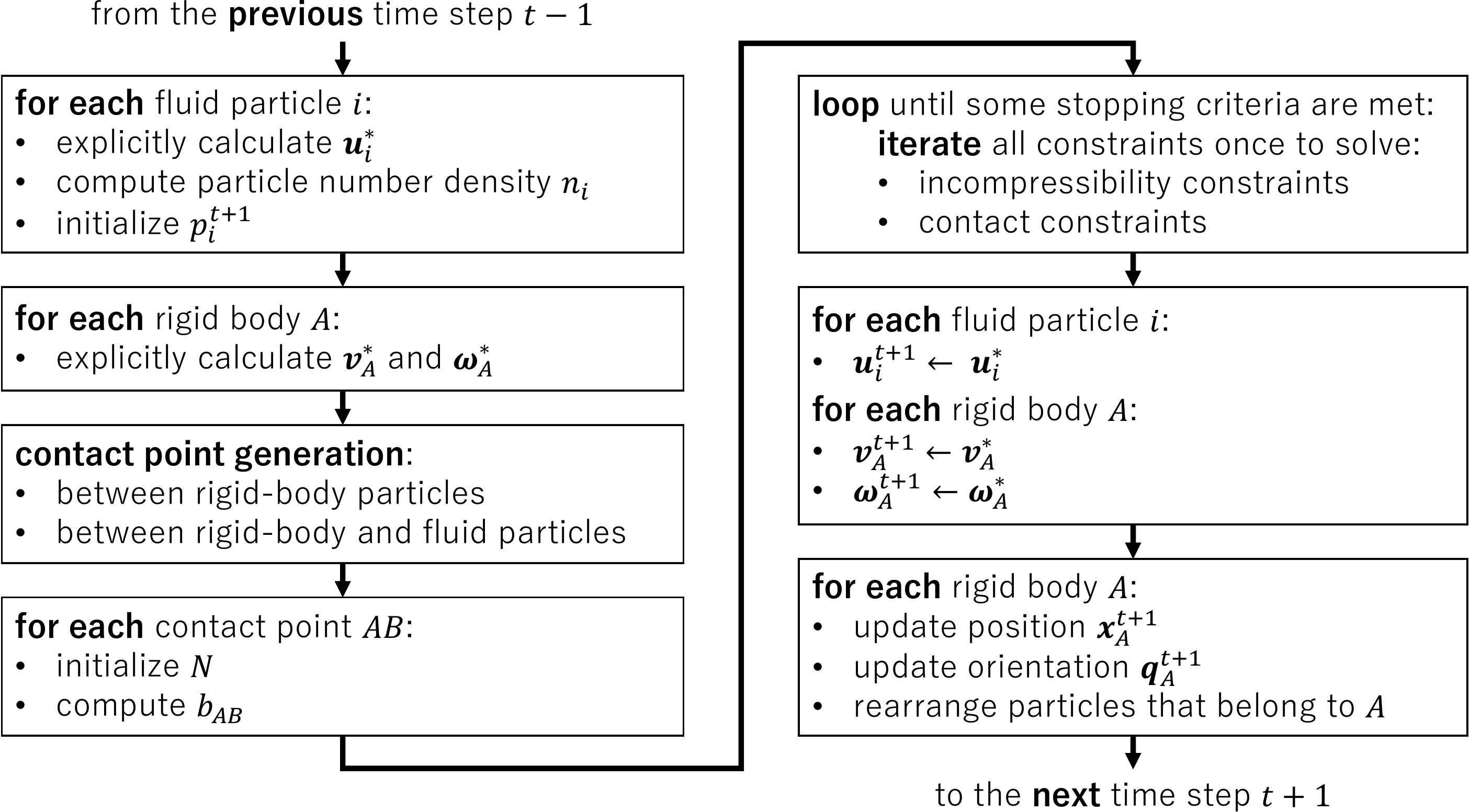}
	\caption{Computation procedure of the entire time step.}
	\label{fig:rigid-fluid:procedure}
\end{figure}

Collision detection and contact generation between rigid bodies and rigid bodies for fluid particles are performed in the same way as those described in the previous subsection. Since every interaction is treated as a rigid-body contact constraint, and a rigid-body contact constraint is a velocity-based constraint, we only need use velocity-based constraints for the entire simulation. This enables us to strongly couple rigid bodies and fluid in the simulation. We finally show the computation procedure of the entire time step for the strongly coupled simulation in \figref{fig:rigid-fluid:procedure}.

\section{Computation Examples}
\label{sec:example}

In this section, we show computation examples of the proposed method. In Section~\ref{sec:example:rigid}, we compute a scene that consists of multiple rigid bodies in order to verify inter-rigid-body collision. In Section~\ref{sec:example:stat}, Section~\ref{sec:example:cpatch}, and Section~\ref{sec:example:dambreak}, we compute scenes that consist of incompressible fluid in order to verify dynamic and static properties of incompressible fluid. In Section~\ref{sec:example:buoyancy} and Section~\ref{sec:example:seesaw}, we compute scenes that consist of both rigid bodies and incompressible fluid in order to verify interaction and strongly-coupled properties between rigid bodies and fluid. Finally, in Section~\ref{sec:example:complex}, we provide a complex scene that involves multiple inter-rigid-body contacts and interaction between rigid bodies and fluid.

\subsection{Rigid-Body Computation}
\label{sec:example:rigid}

\begin{figure}
	\centering
	\includegraphics[width=0.5\textwidth]{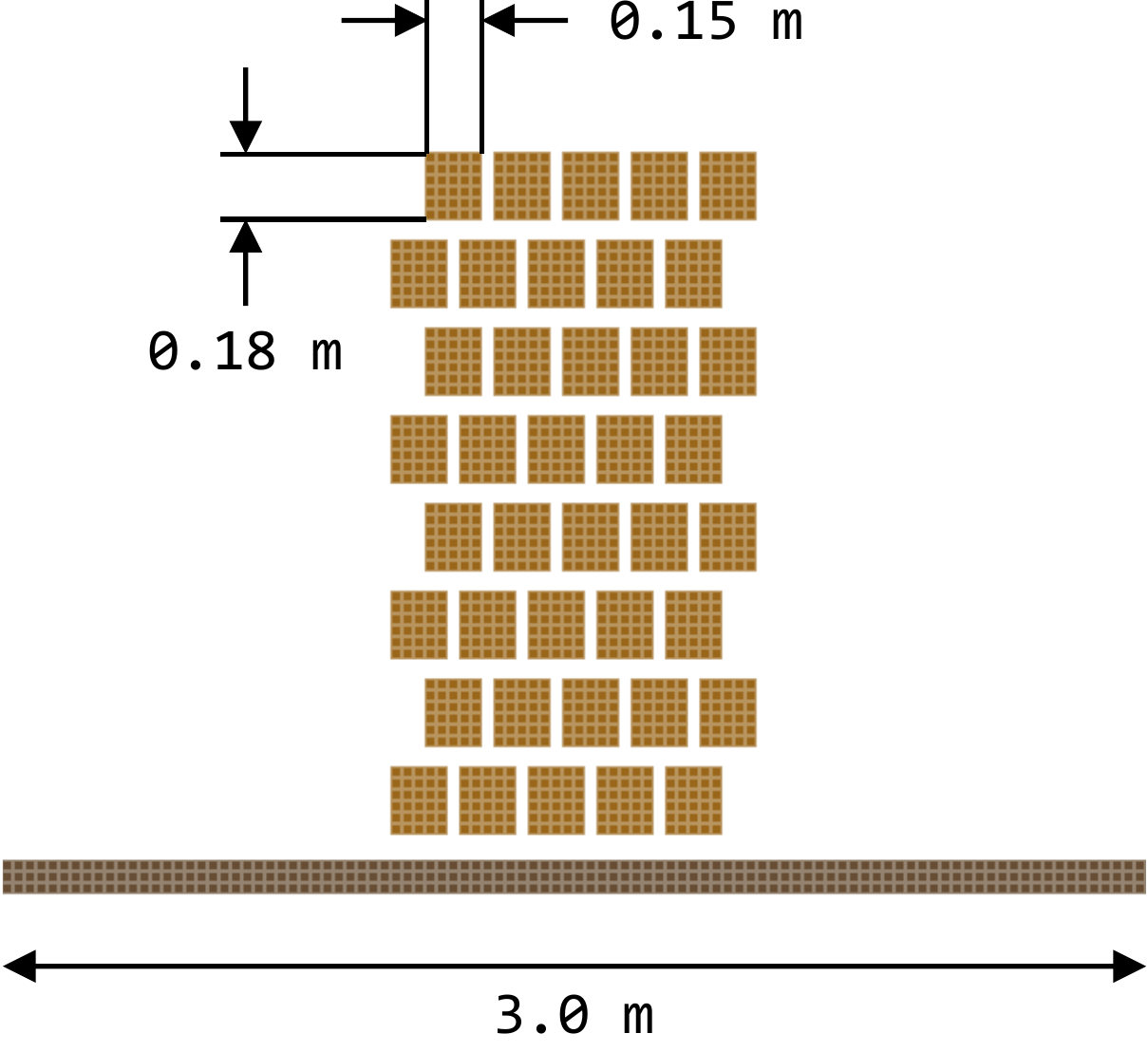}
	\caption{Initial configuration of the rigid-body simulation.}
	\label{fig:example:rigid-init}
\end{figure}

\begin{figure}
	\centering
	\includegraphics[width=0.6\textwidth]{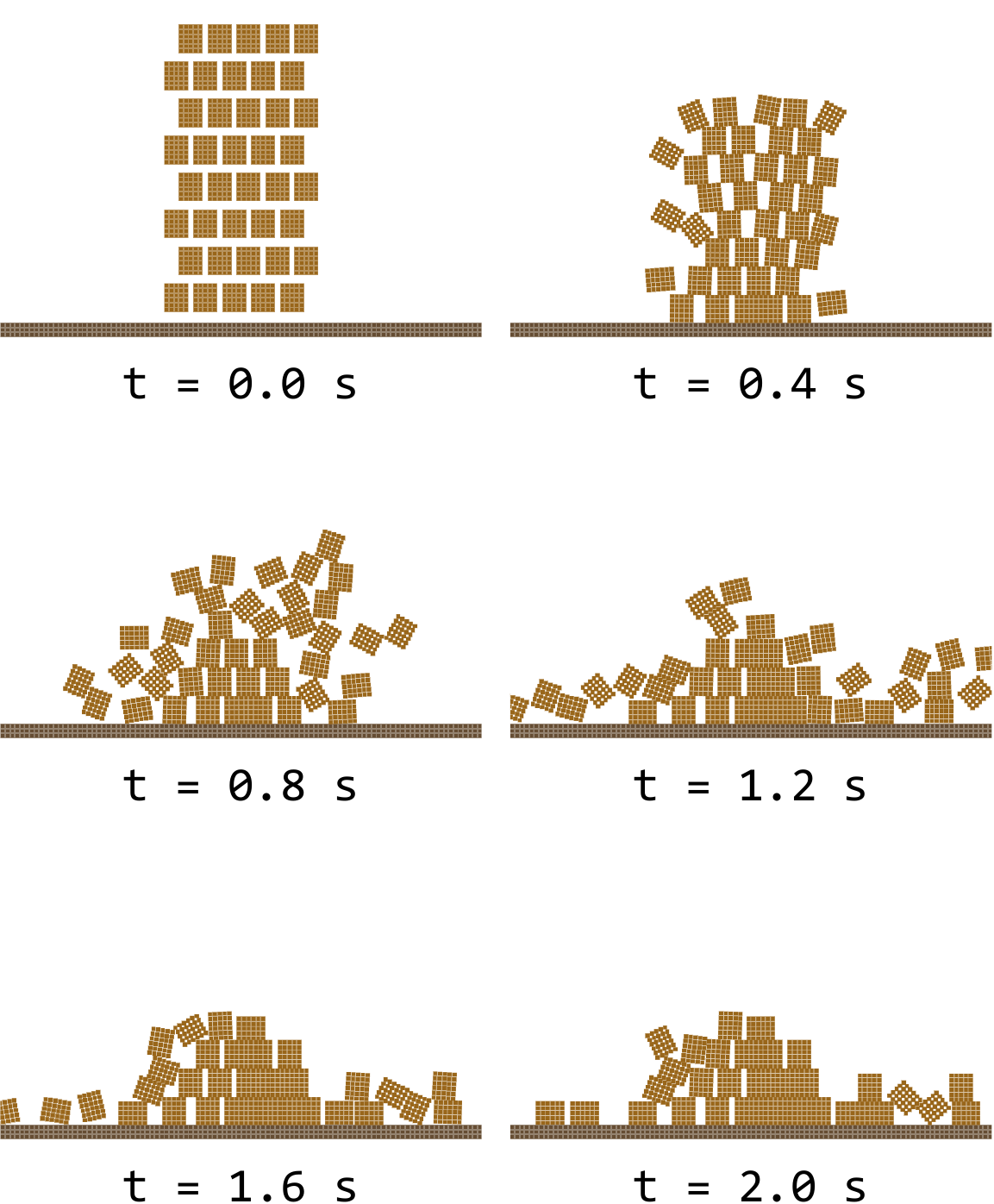}
	\caption{Snapshots of the rigid-body simulation at times 0.0, 0.4, 0.8, 1.2, 1.6, and 2.0 in seconds.}
	\label{fig:example:rigid}
\end{figure}

We show a two-dimensional computation example of rigid-body interaction we introduced in Section \ref{sec:rigid}. The initial configuration of the simulation is shown in \figref{fig:example:rigid-init}. Each box has the same mass, and the floor below the boxes has infinite mass and is not affected by gravity. The coefficient of restitution is set to 0.2 for all contact points. The interval of the particle arrangement $l$ is equal to 0.03 m, and the time step $h$ is 0.005 s. The gravitational acceleration $g$ is set to 9.8 m/s$^2$ Snapshots of the simulation are shown in \figref{fig:example:rigid}. The piled boxes are dropped and collapsed, scattering over the floor. Thanks to the mixed term of constraint \eqref{eq:rigid:mixed}, no penetration between rigid bodies can be found in \figref{fig:example:rigid}. Although no friction is considered, we can see the friction-like effect in the simulation since the surfaces of the rigid bodies consist of particles and are uneven.

\subsection{Hydrostatic Pressure Computation}
\label{sec:example:stat}

\begin{figure}
	\centering
	\includegraphics[width=0.5\textwidth]{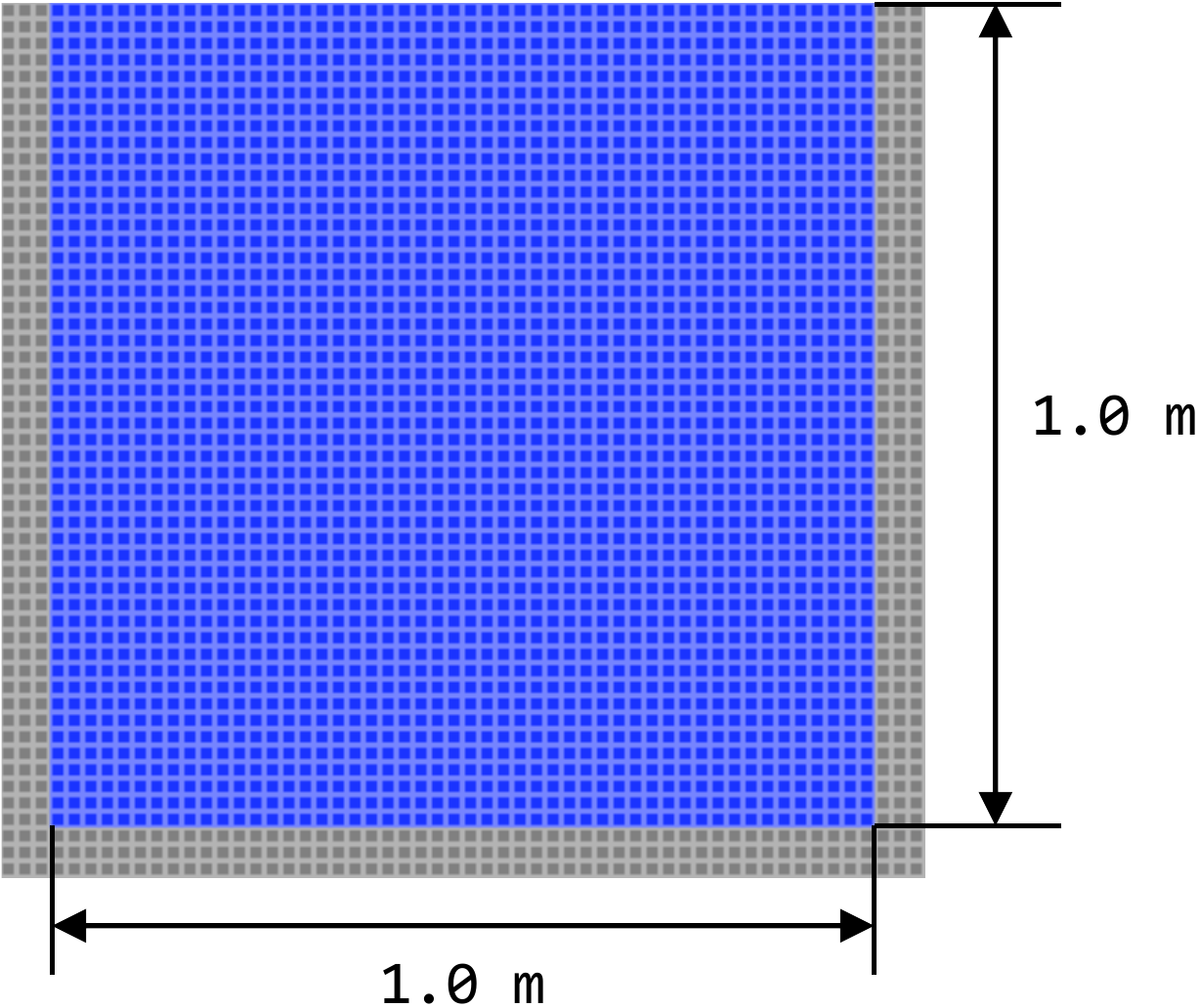}
	\caption{Initial configuration of the hydrostatic pressure computation.}
	\label{fig:example:static-init}
\end{figure}

\begin{figure}
	\centering
	\includegraphics[width=0.8\textwidth]{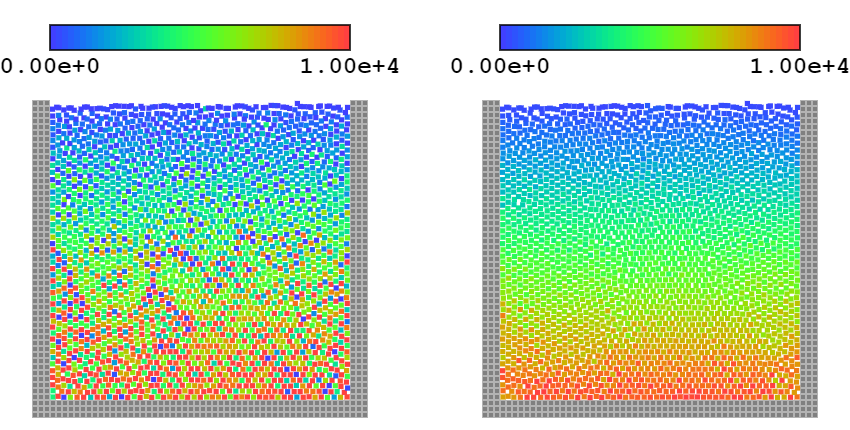}
	\caption{Raw (left) and smoothed (right) pressure distribution at time 1.0 s. The values are shown in pascals.}
	\label{fig:example:static}
\end{figure}

\begin{figure}
	\centering
	\includegraphics[width=0.5\textwidth]{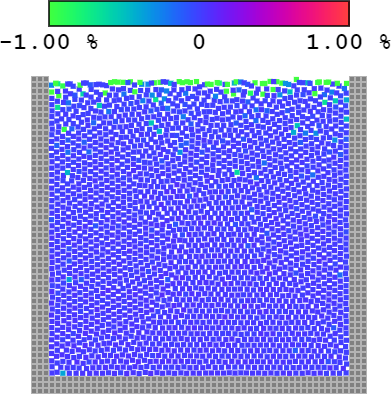}
	\caption{Distribution of the deviation from the standard particle number density at time 1.0 s.}
	\label{fig:example:static-pnd}
\end{figure}

Since we introduce a novel method for fluid computation, it is worthwhile to confirm the basic properties of the fluid. We compute the hydrostatic pressure of fluid. The initial configuration is shown in \figref{fig:example:static-init}. We set $l$ to 0.02 m, $r_e/l$ to 2.1, $g$ to 10.0 m/s$^2$, and $h$ to 0.001 s. We show the distributions of the raw pressure $p$ and the smoothed pressure $\tilde p$ that is computed according to \eqref{eq:fluid:spressure} at time 1.0 s in \figref{fig:example:static}. From \figref{fig:example:static}, we can observe that the raw pressure distribution suffers from heavy noise, whereas the smoothed distribution appears better and to be accurate. As we describe in Section \ref{sec:fluid:weighting}, the smoothing length is equal to $r_e$, and thus the noise can be said to have only a local and limited effect. In addition, the distribution of the deviation from the standard particle number density $(n_i-n_0)/n_0$ for each particle $i$ at time 1.0 s is shown in \figref{fig:example:static-pnd}. From the figure, we can see that no particle is compressed to the level of 1.0\% deviation, and the incompressibility is highly assured.

\subsection{Dam-break Computation}
\label{sec:example:dambreak}

\begin{figure}
	\centering
	\includegraphics[width=0.5\textwidth]{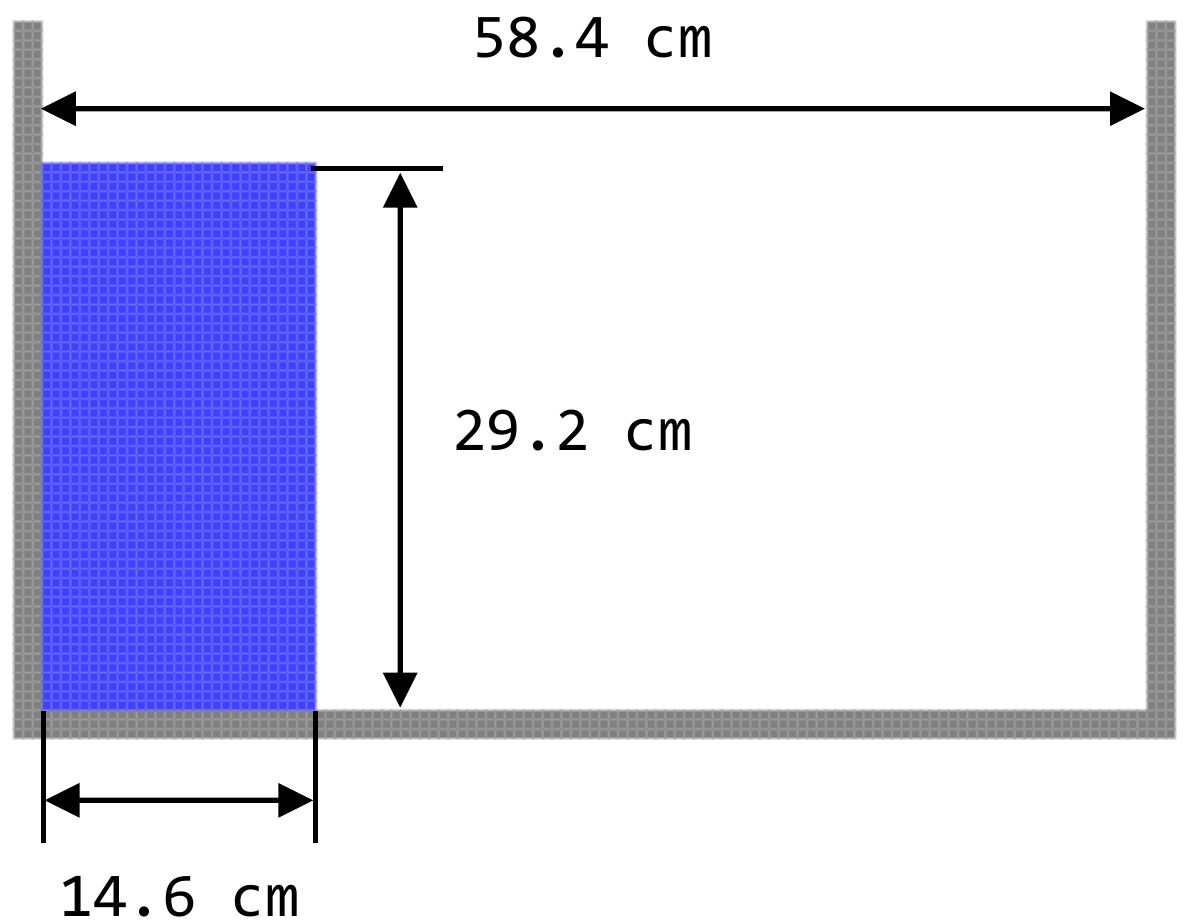}
	\caption{Initial configuration of the dam break computation.}
	\label{fig:example:dambreak-init}
\end{figure}

\begin{figure}
	\centering
	\includegraphics[width=0.65\textwidth]{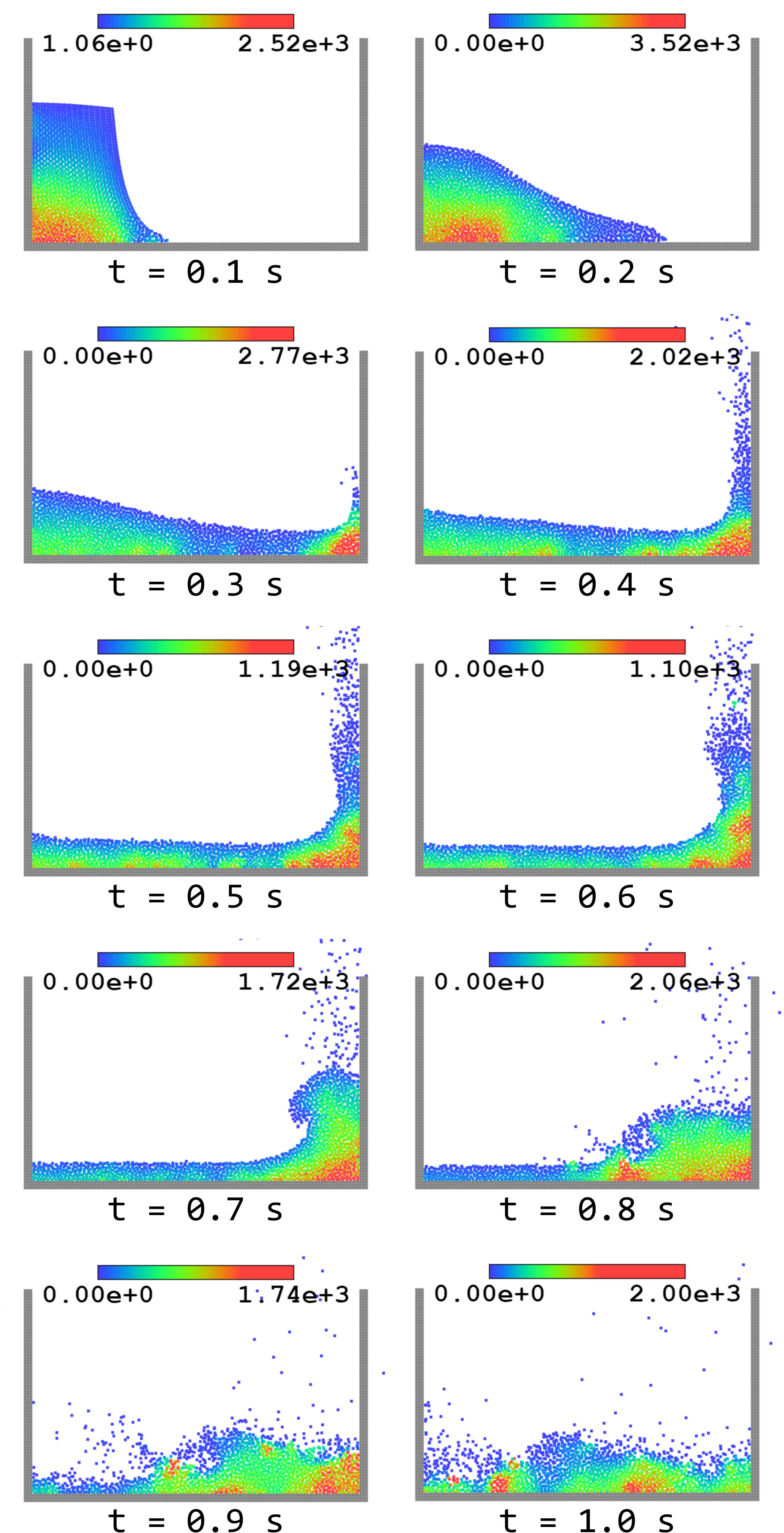}
	\caption{Snapshots of the dam break computation at times from 0.1 s to 1.0 s, in steps of 0.1 s. Each fluid particle is colored based on its smoothed pressure, and the numbers shown at the top of each figure represent the minimum and maximum smoothed pressures in pascals.}
	\label{fig:example:dambreak}
\end{figure}

The dam-break problem is simulated using the proposed method. The initial configuration is shown in \figref{fig:example:dambreak-init}. This is the same configuration as in the experiment performed by Koshizuka and Oka~\cite{koshizuka1996moving}. We set $l$ to 0.005 m, $r_e/l$ to 2.1, $g$ to 9.80665 m/s$^2$, and $h$ to 0.0005 s. We show the snapshots of the simulation in \figref{fig:example:dambreak}, which is colored according to smoothed pressure. We can observe that the fluid reaches the right wall approximately $t$ = 0.3 s after the beginning of the simulation and causes a heavy splash. The fluid makes a breaking wave from approximately $t$ = 0.7 s to $t$ = 0.9 s. The features of the simulation visibly match those of the experiment.

\subsection{Circular Patch Computation}
\label{sec:example:cpatch}

\begin{figure}
	\centering
	\includegraphics[width=0.9\textwidth]{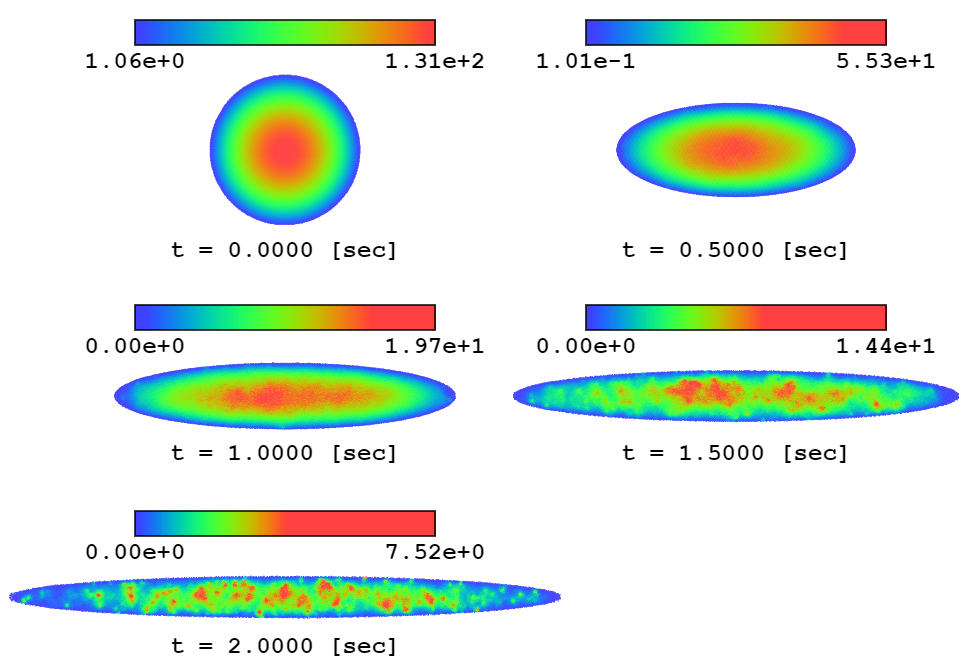}
	\caption{Snapshots of the circular patch computation at times 0.0, 0.5, 1.0, 1.5, and 2.0 in seconds. The numbers at the top of each snapshot indicate the pressure in pascals.}
	\label{fig:example:cpatch}
\end{figure}

\begin{figure}
	\centering
	\begin{minipage}{0.6\textwidth}
		\centering
		\includegraphics[width=\hsize]{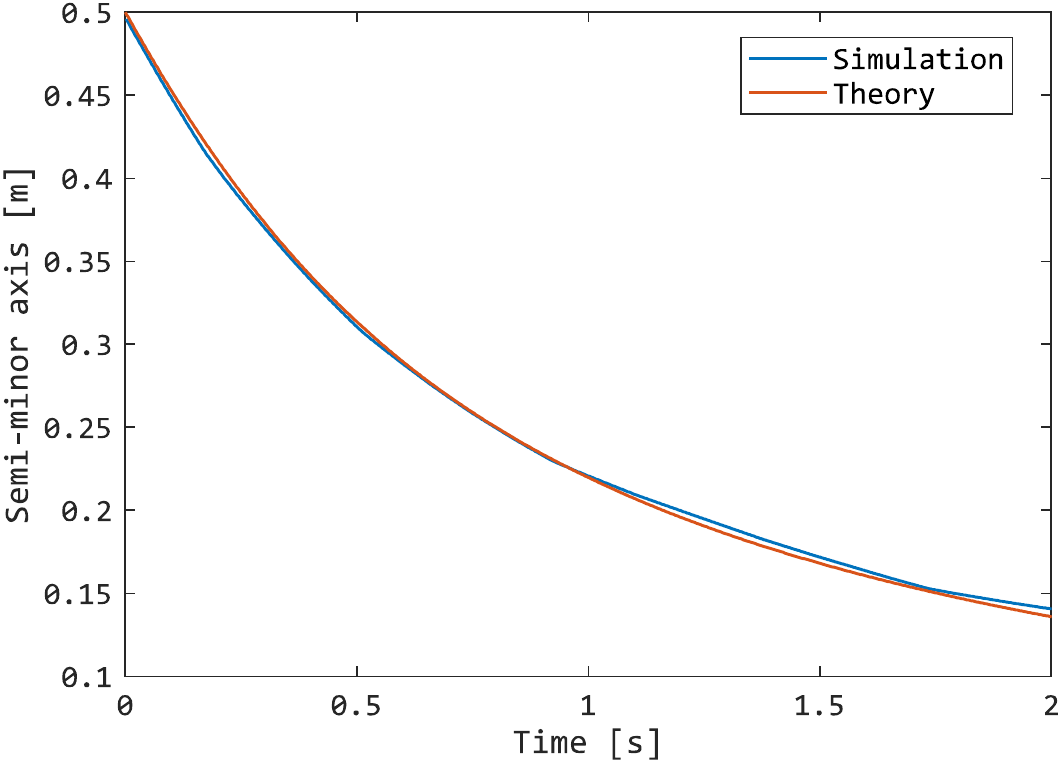}
		\subcaption{Semi-minor axis}
		\vspace{0.3cm}
	\end{minipage}
	\begin{minipage}{0.6\textwidth}
		\includegraphics[width=\hsize]{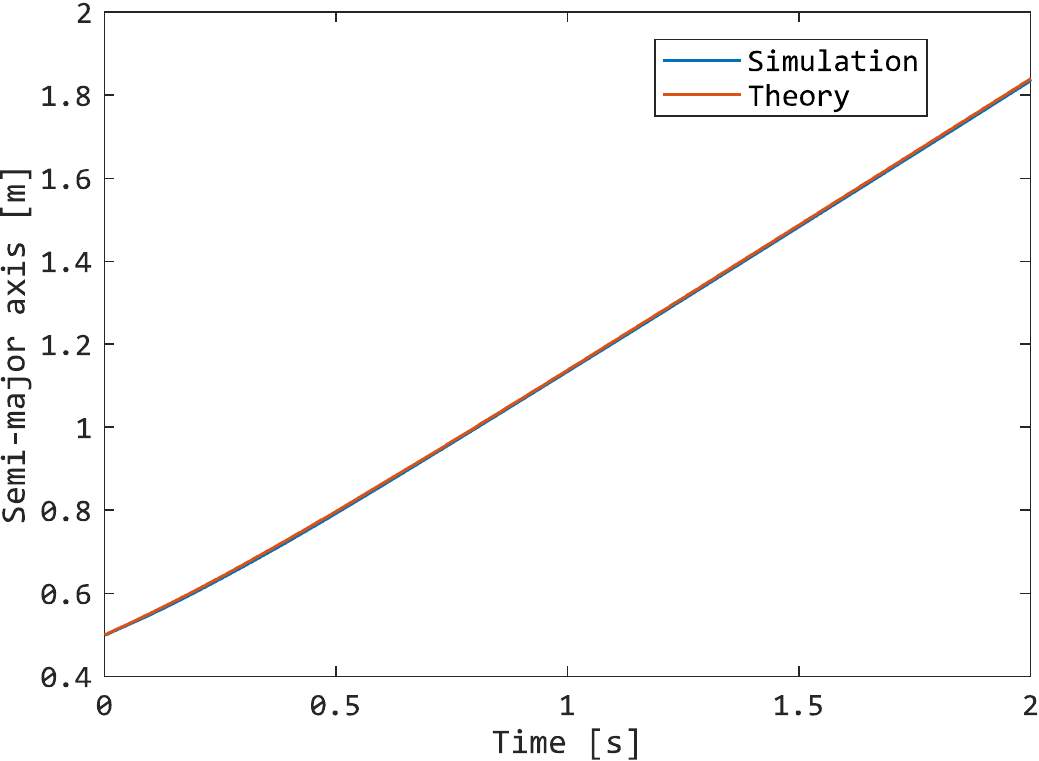}
		\subcaption{Semi-major axis}
	\end{minipage}
	\caption{The result of the circular patch computation. The horizontal axes show the time, and the vertical axes show the time evolution of the semi-minor axis (a) and the semi-major axis (b).}
	\label{fig:example:cpatch-axes}
\end{figure}

An extending circular droplet is computed using the proposed method. The radius of the droplet is set to 0.5 m, and the center of the droplet is located at the origin. The initial velocity of each particle $i$ at $(x,y)$ is given by
\begin{align}
	\u_i = \begin{pmatrix}
		x \\
		-y
	\end{pmatrix}.
\end{align}
We set $l$ to 0.005 m, $r_e/l$ to 2.5, $g$ to 0 m/s$^2$, and $h$ to 0.002 s. This problem can be solved analytically~\cite{colagrossi2005meshless} and the theoretical solution is given by
\begin{align}
	\frac{da}{dt}     & = -aA                                            \\
	\frac{db}{dt}     & = bA                                             \\
	\frac{d^2A}{dt^2} & = \frac{4}{A}\left(\frac{dA}{dt}\right)^2 - 2A^3
\end{align}
with boundary conditions
\begin{align}
	a(0) = 0.5,\ b(0) = 0.5,\ \frac{dA}{dt}(0) = 0,\ A(0) = 1,
\end{align}
where $a$ and $b$ are the semi-minor axis and the semi-major axis of the droplet in meters, respectively. Snapshots are shown in \figref{fig:example:cpatch}, and the simulation result is shown in \figref{fig:example:cpatch-axes}. The theoretical result in \figref{fig:example:cpatch-axes} is computed numerically. Although pressure disturbance can be observed as the droplet extends due to the disorder of the particle arrangement, the graph shows the simulated values of the semi-minor axis and the semi-major axis well match the theoretical result.

\subsection{Buoyancy Computation}
\label{sec:example:buoyancy}

\begin{figure}
	\centering
	\includegraphics[width=0.6\textwidth]{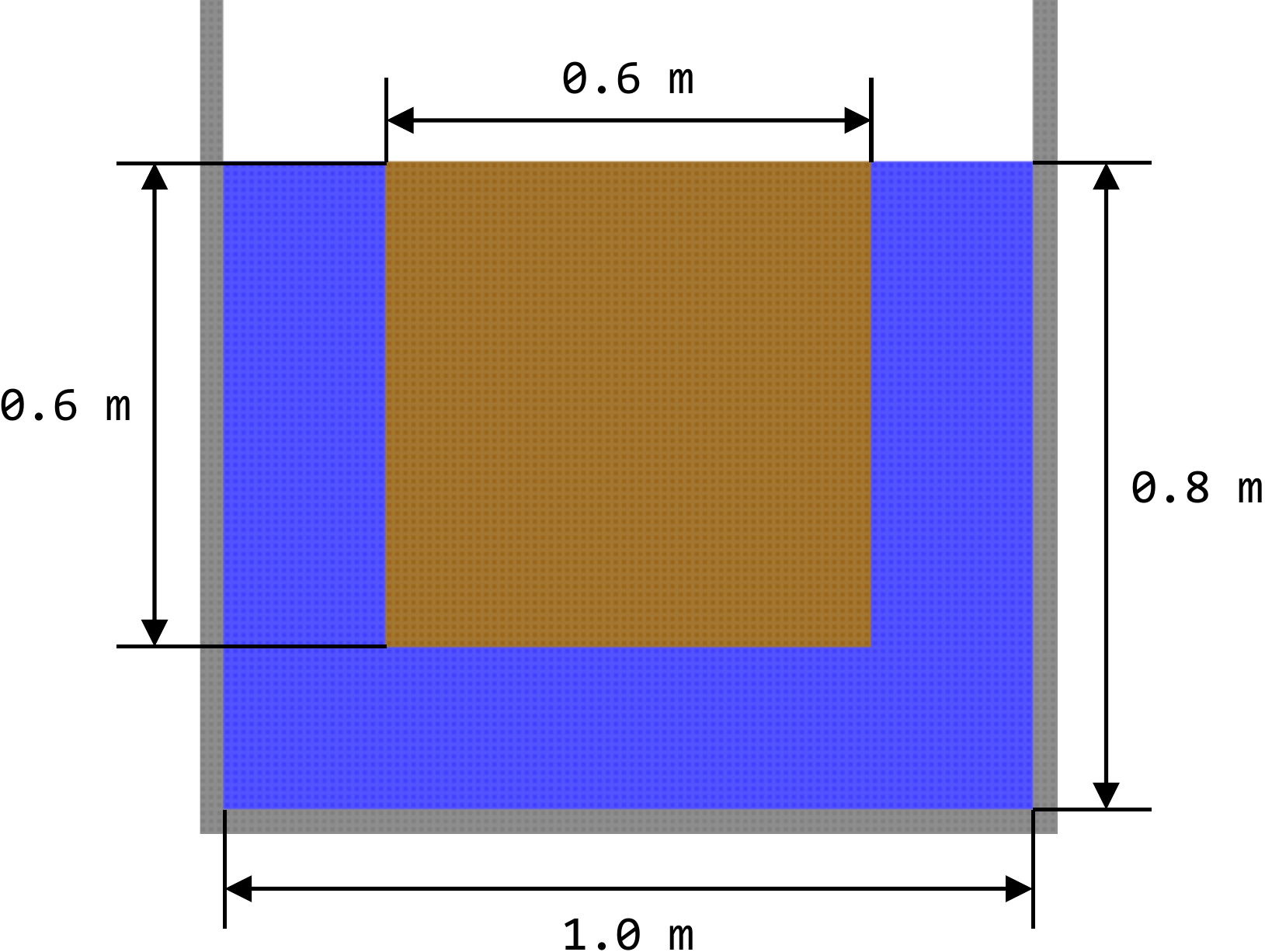}
	\caption{Initial configuration of the buoyancy computation.}
	\label{fig:example:buoyancy-init}
\end{figure}

\begin{figure}
	\centering
	\includegraphics[width=0.6\textwidth]{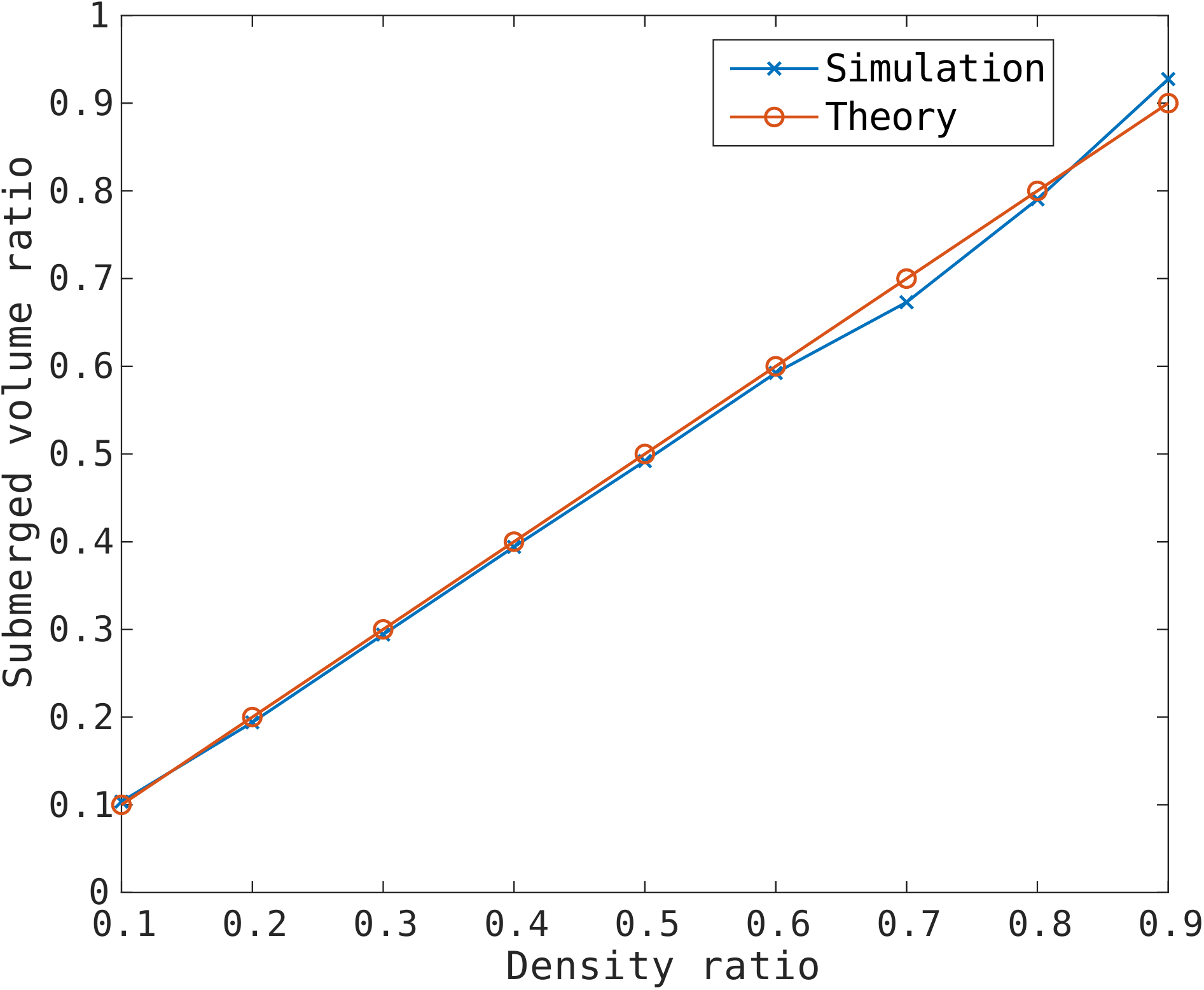}
	\caption{Result of the buoyancy computation. The horizontal axis shows the density of the box relative to the density of the fluid, and the vertical axis shows the ratio of the submerged volume of the box.}
	\label{fig:example:buoyancy}
\end{figure}

In order to confirm that the interaction between a rigid body and fluid is computed correctly, we put boxes with different densities into a fluid to observe the effect of buoyancy. The initial configuration of the simulation is shown in \figref{fig:example:buoyancy-init}. The box is a square of edge length 0.6 m, and its density is set to have a specified density ratio to the fluid. The density ratio varies from 0.1 to 0.9 in steps of 0.1. We set $l$ to 0.01 m, $r_e/l$ to 2.1, $g$ to 9.80665 m/s$^2$, and $h$ to 0.0025 s. We measure the ratio of the submerged volume of the box when the box becomes stationary in the fluid, and the result is shown in \figref{fig:example:buoyancy}. The graph shows that the simulated values match the theoretical values well.

\subsection{Seesaw Computation}
\label{sec:example:seesaw}

\begin{figure}
	\centering
	\includegraphics[width=0.4\textwidth]{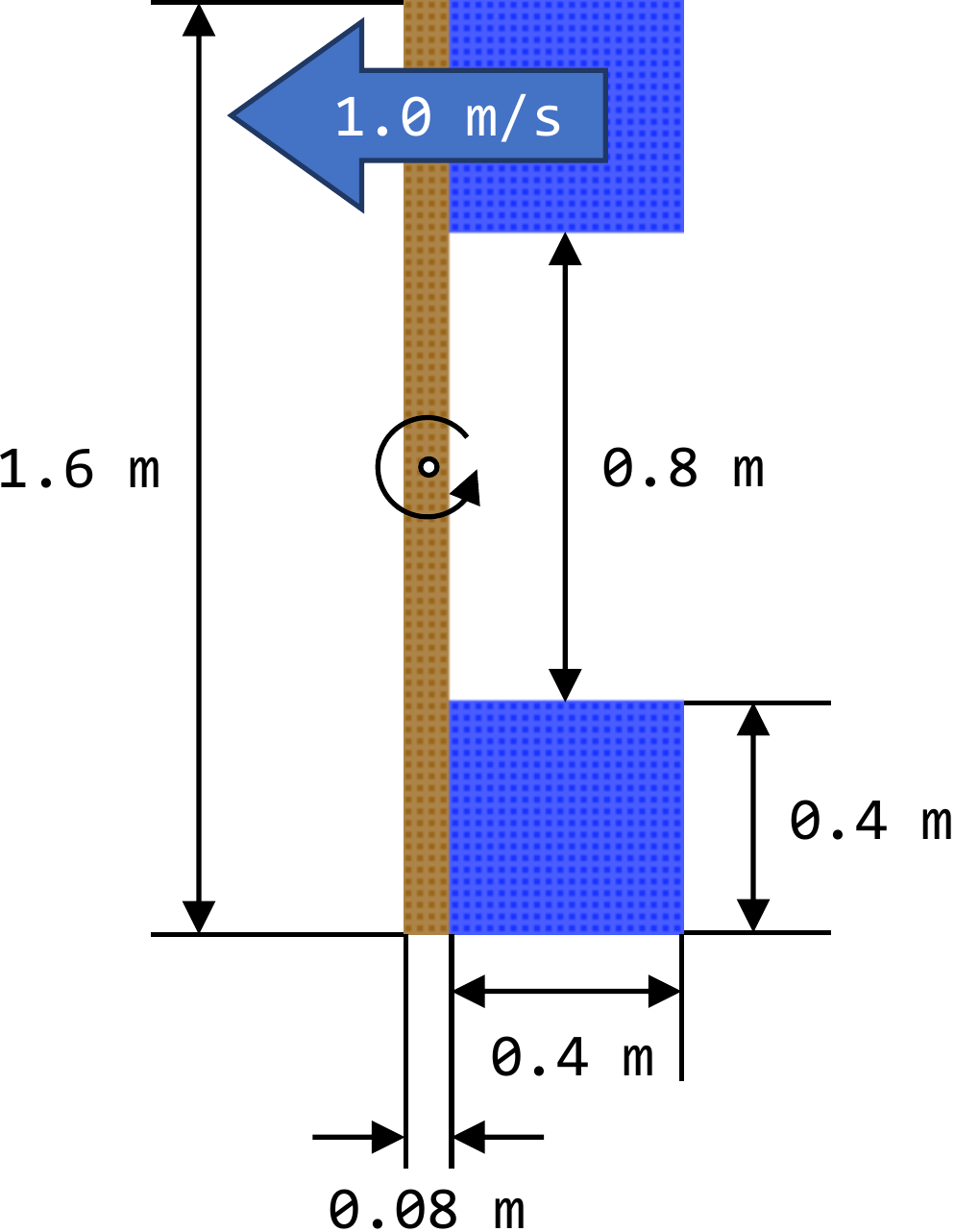}
	\caption{Initial configuration of the seesaw computation.}
	\label{fig:example:seesaw-init}
\end{figure}

\begin{figure}
	\centering
	\includegraphics[width=0.6\textwidth]{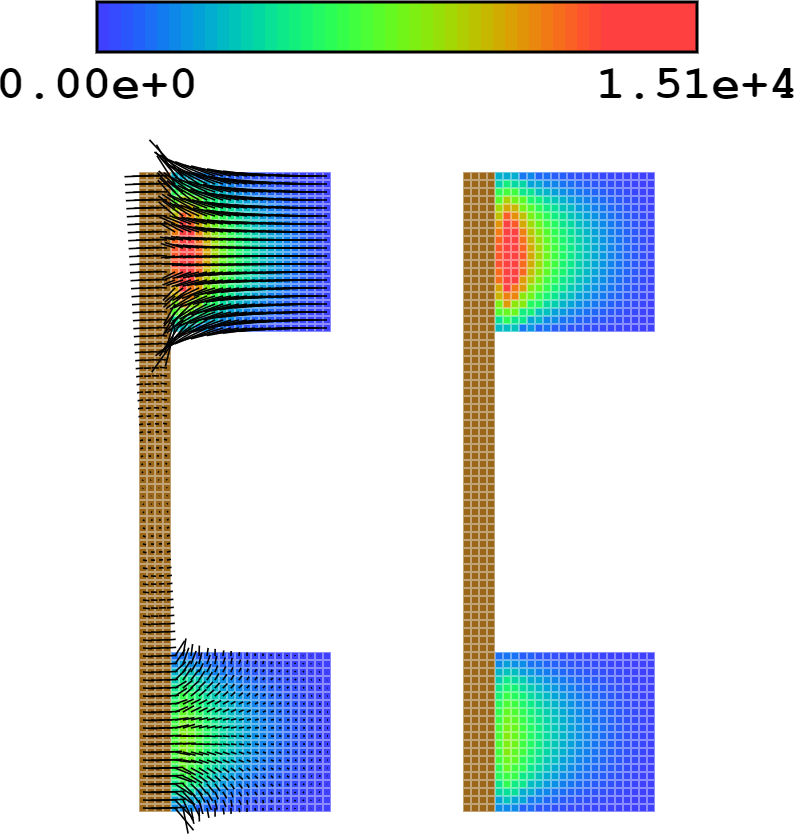}
	\caption{Pressure distribution with (left) and without (right) velocity distribution just after all constraints are solved. The numbers at the top indicate the pressure in pascals.}
	\label{fig:example:seesaw}
\end{figure}

\begin{figure}
	\centering
	\includegraphics[width=0.6\textwidth]{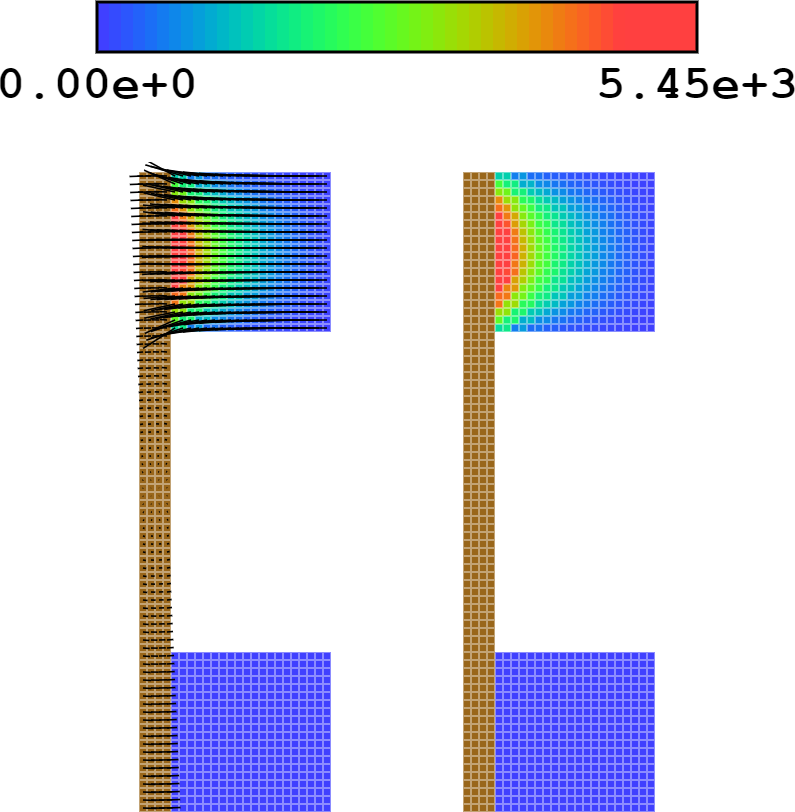}
	\caption{Pressure distribution with (left) and without (right) velocity distribution just after updating the velocity of the rigid body and the fluid. The interaction of the rigid body and the fluid is computed using the passively moving solid model. The numbers at the top indicate the pressure in pascals.}
	\label{fig:example:seesaw-pms}
\end{figure}

In strongly coupled simulations, multiple interactions of rigid bodies and fluid can be computed simultaneously. To confirm this, we run the following seesaw computation. The initial configuration of the simulation is shown in \figref{fig:example:seesaw-init}. The densities of the rigid body and fluid are 490 kg/m$^2$ and 1,000 kg/m$^2$, respectively. The top square part of the fluid has an initial velocity of 1.0 m/s toward the left, and other fluid and the rigid body are set to be still. The rigid body is pinned at and can only rotate around its center of gravity. We set $l$ to 0.02 m, $r_e/l$ to 2.1, $g$ to 0.0 m/s$^2$, and $h$ to 0.005 s. If the simulation is strongly coupled, all of the following things happen instantly in a single time step. The top part of the fluid pushes the rigid body toward the left so that the rigid body starts to rotate and pushes the bottom part of the fluid toward the right. This results in positive pressure at the bottom part of the fluid. \figref{fig:example:seesaw} shows the pressure distribution and the velocity distribution just after solving all constraints and before updating positions. We can observe that the rigid body is rotating and the bottom part of the fluid has positive pressure due to the rotation. We also show the result obtained with the PMS model in \figref{fig:example:seesaw-pms}. Since the PMS model cannot handle strong coupling, the bottom part of the fluid has no positive pressure and zero velocity. In subsequent steps, however, the PMS model can handle weakly-coupled interaction between the rigid body and the lower part of the fluid and positive pressure will be observed there. Note that the result obtained with the PMS model shows a lower peak pressure, as compared to the proposed method, because the top part of the fluid cannot take into account the existence of the bottom part of the fluid. As such, it is easier for the top part of the fluid to push the rigid body.

\subsection{Complex Scene}
\label{sec:example:complex}

\begin{figure}
	\centering
	\includegraphics[width=0.8\textwidth]{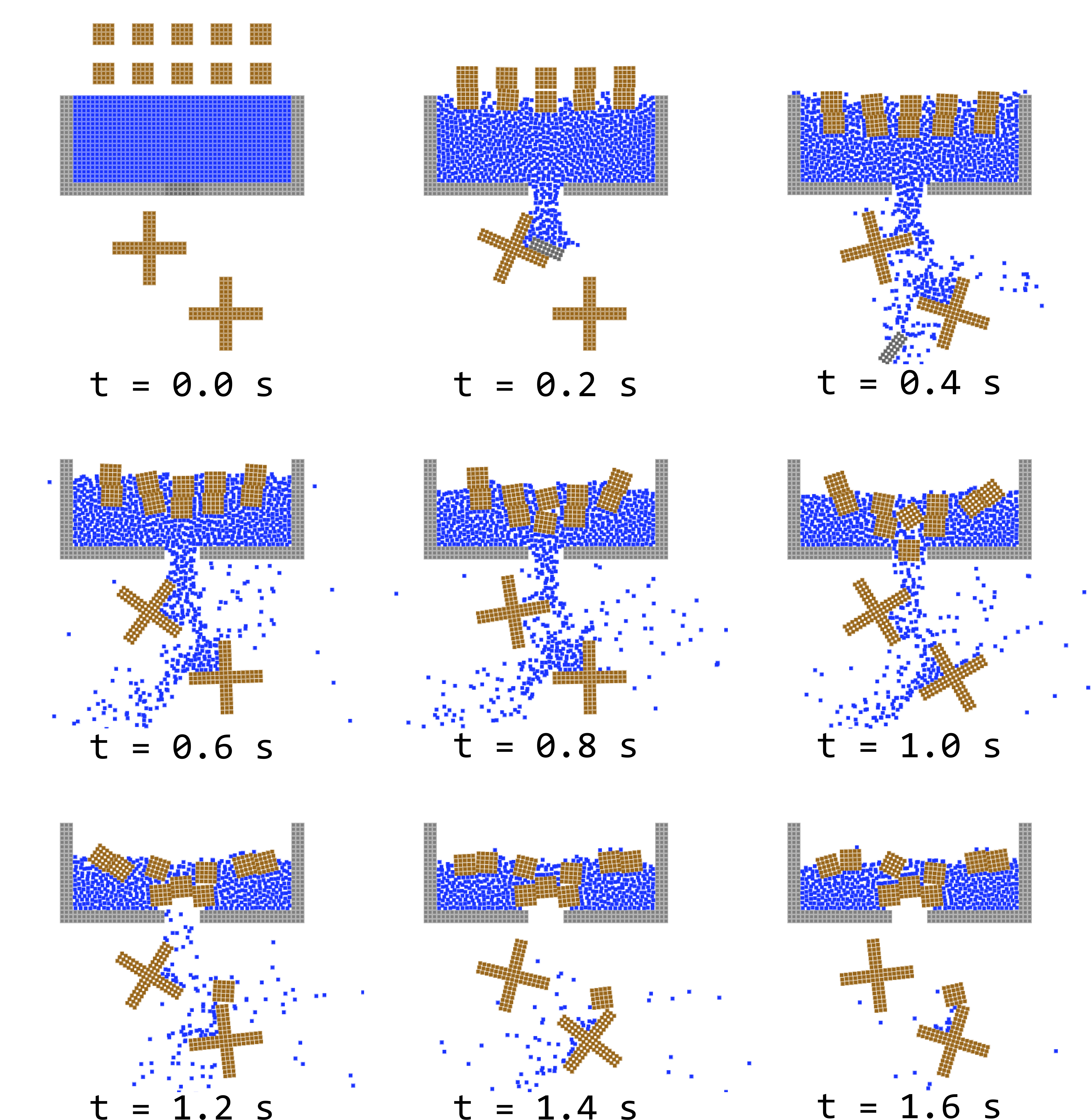}
	\caption{Snapshots of the complex simulation at times from 0.0 s to 1.6 s.}
	\label{fig:example:complex}
\end{figure}

Finally, we give a computation example of a complex and dynamic scene that contains both inter-rigid-body collision and rigid bodies and fluid interaction. The centers of gravity of two cross-shaped rigid bodies at the bottom are fixed and can only rotate around the centers of gravity. The other rigid bodies are not fixed and can move freely. The simulation result is shown in \figref{fig:example:complex}. We set $l$ to 0.02 m, $r_e/l$ to 2.1, $g$ to 9.80665 m/s$^2$, and $h$ to 0.001 s. We can observe that the simulation visibly runs without problems.

\section{Conclusion}
\label{sec:conclusion}

In this research, we proposed a method to simulate an incompressible fluid that uses an LCP to formulate the incompressibility constraint and enables strong coupling with rigid bodies. We formulated velocity-based constraints that generalize incompressibility constraints in fluid computation and non-penetration constraints in rigid-body computation, which provides the general framework of various strongly coupled simulations. Through numerical examples, we have demonstrated that the proposed method can compute incompressible fluid accurately and achieves strong coupling with rigid bodies correctly. With this method, we can use ordinary impulse-based methods for rigid-body simulation; therefore, the proposed method fits in well with existing software for rigid-body simulation.

The remaining problems of the proposed method include the fact that it is difficult to compute negative pressure because we required the pressure to be nonnegative in order not to cause an attractive force between particles. Some stabilization technique is likely needed to allow for negative pressure. In addition, generalization of the shape representation of rigid bodies is desirable so that we can represent a smoother face and decrease the computation time by means other than the use of particles. We leave these as future research topics.

\end{document}